\documentclass[acmsmall, review=false, authorversion, nonacm]{acmart}

\usepackage{algorithmicx, algorithm}
\usepackage{algpseudocode}
\usepackage{mathtools}
\usepackage{physics}
\usepackage{qcircuit}
\usepackage{xcolor}
\usepackage{subcaption,xspace}
\usepackage{thmtools,thm-restate}
\usepackage{bbm}

\newcommand{\st}[1]{}
\newcommand{\xcancel}[1]{}

\newcommand{\E}{\mathbb{E}}
\newcommand{\half}{\frac{1}{2}}
\newcommand{\thalf}{\tfrac{1}{2}}
\newcommand{\xor}{\oplus}

\newcommand{\nlf}{\eta(f)}
\newcommand{\fmax}{\hat{f}_{\max}}

\newcommand{\Ftwo}[1]{\{0,1\}}

\newcommand{\Ot}{\tilde{O}}
\newcommand{\fcp}{\mathfrak{cp}}
\newcommand{\fp}{\mathfrak{p}}
\newcommand{\pwc}{PWC\xspace}
\newcommand{\bfmax}{{\tt BoundFMax} }
\newcommand{\pwce}{PWCE }

\newcommand{\pmset}{S}
\newcommand{\hdq}[1]{{\tt HD_{#1}}}
\newcommand{\cmp}{{\tt CMP}}
\newcommand{\cuquad}{\qquad\qquad\qquad\qquad\quad}
\newcommand{\hatf}[1]{\hat{f}(#1)}
\newcommand{\bre}[1]{\Breve{#1}}
\newcommand{\eqae}{EQAmpEst\xspace}

\declaretheorem[name=Proposition, numberwithin=section]{proposition}
\DeclarePairedDelimiter{\ceil}{\lceil}{\rceil}

\allowdisplaybreaks

\algnewcommand\algorithmicon{\textbf{on}}
\algdef{SE}[ON]{On}{EndOn}[1]{\algorithmicon\ #1\ \algorithmicdo}{\algorithmicend\ \algorithmicloop}

\AtBeginDocument{%
  \providecommand\BibTeX{{%
    \normalfont B\kern-0.5em{\scshape i\kern-0.25em b}\kern-0.8em\TeX}}}

\setcopyright{acmcopyright}
\copyrightyear{2018}
\acmYear{2018}
\acmDOI{10.1145/1122445.1122456}

\begin{document}

\title{Quantum and Randomised Algorithms for Non-linearity Estimation}

\author{Debajyoti Bera}
\email{dbera@iiitd.ac.in}
\affiliation{%
  \institution{IIIT-Delhi}
  \city{New Delhi}
  \state{Delhi}
  \postcode{110020}
}
\author{SAPV Tharrmashastha}
\email{tharrmashasthav@iiitd.ac.in}
\affiliation{%
  \institution{IIIT-Delhi}
  \city{New Delhi}
  \state{Delhi}
  \postcode{110020}
}

\renewcommand{\shortauthors}{D. Bera and S. Tharrmashastha}

\begin{abstract}
    Non-linearity of a Boolean function indicates how far it is from any linear function. Despite there being several strong results about identifying a linear function and distinguishing one from a sufficiently non-linear function, we found a surprising lack of work on computing the non-linearity of a function. The non-linearity is related to the Walsh coefficient with the largest absolute value; however, the naive attempt of picking the maximum after constructing a Walsh spectrum requires $\Theta(2^n)$ queries to an $n$-bit function. We improve the scenario by designing highly efficient quantum and randomised algorithms to approximate the non-linearity allowing additive error, denoted $\lambda$, with query complexities that depend polynomially on $\lambda$. We prove lower bounds to show that these are not very far from the optimal ones. The number of queries made by our randomised algorithm is linear in $n$, already an exponential improvement, and the number of queries made by our quantum algorithm is surprisingly independent of $n$. Our randomised algorithm uses a Goldreich-Levin style of navigating all Walsh coefficients and our quantum algorithm uses a clever combination of Deutsch-Jozsa, amplitude amplification and amplitude estimation to improve upon the existing quantum versions of the Goldreich-Levin technique.

\end{abstract}


\maketitle

\section{Introduction}\label{sec:intro}

Boolean functions are an indispensable tool to design ciphers, codes
and algorithms. Of particular interest are ``simple'' Boolean functions and the ``hard'' ones. The most common candidates are linear and bent functions, respectively, and the characterisation used for them is their {\em non-linearity}. Non-linearity of a function is 
defined as the smallest (Hamming) distance of that function to {\em any} affine
function and is one of the conceptually simplest metric to evaluate a Boolean function.

Our subjects of investigation are $n$-bit input one-bit output Boolean functions. For any
two such functions $f,g:\{0,1\}^n \to \{0,1\}$, define $dist(f,g)$ as the Hamming
distance between the truth-tables of $f$ and $g$. For any $a \in \{0,1\}^n$,
define the function $\chi_a:\{0,1\}^n \to \{0,1\}$ as $\chi_a(x) = {x \cdot a} \text{ (mod 2)}$
in which $x \cdot a$ denotes the binary bitwise dot-product between $a$ and $x$.
It turns out that these functions $\{\chi_a(\cdot)\}_{a \in \{0,1\}^n}$ exactly
represent the set of $n$-bit linear functions --- these are the functions for
which $g(x \xor y) = g(x) \xor g(y)$. Affine Boolean functions (those with
algebraic degree 1) are generalisations of linear functions and satisfy $g(x \xor y) = g(x) \xor g(y) \xor b$ for $b \in \{0,1\}$; they are represented by functions
$\chi'_{a,b}(x) = (x \cdot a)\xor b$. Of course, $\chi_a = \chi'_{a,0}$.

The Walsh coefficient of a function $f$ at a point $a$ is defined as the average correlation of $f$ with $\chi_a$. We normalise it slightly differently to suit our approaches but that does not affect the main results of this work.
We use the following definition.
$$\hat{f}(a) = \frac{1}{2^n} \sum_x (-1)^{f(x) \xor \chi_a(x)} = \frac{1}{2^n} \sum_x (-1)^{f(x) \xor x \cdot a}.$$

Any Boolean function whose Walsh coefficients have the same absolute value is called a {\em bent function}; we know from Parseval's inequality that these functions satisfy $\hat{f}(a) = \pm\frac{1}{\sqrt{2^n}}$ for all $a\in \{0,1\}^n$. Such functions are considered to be the most ``non-linear'' ones. On the other hand, linear functions satisfy $\hat{f}(a)=1$ only when $f=\chi_a$, and $\hat{f}(a)=0$ when $f$ is some different linear function. Observe that $\max_{a} \{|\hat{f}(a)|\}$ (which we denote $\fmax$) satisfies $\frac{1}{\sqrt{2^n}} \le \fmax \le 1$ --- the first equality holds for Bent functions and the second equality holds for linear functions.

The (normalised) non-linearity of any $n$-bit Boolean function $f$ is defined as
the minimum Hamming distance from $f$ to any affine Boolean function; mathematically,
$$\nlf = \min_{a,b} \frac{dist(f, \chi'_{a,b})}{2^n}. $$
Using the above notation $0 \le
\nlf \le \thalf - \thalf \cdot \frac{1}{2^{n/2}}$; however, this definition is computationally expensive to operate since one has to
enumerate over all possible (exponentially many) linear functions and then
compute distance between those functions and $f$ that also requires
exponentially many function evaluations, giving a running time of $\Theta(2^{2n})$.

A seemingly simpler alternative arises from the Walsh-Hadamard (WH) transform of
$f$. The WH transform generates a $2^n$-dimensional WH spectrum whose $a$-th
coefficient is $\hat{f}(a)$.
It is easy to see that $\hat{f}(a)$ can be expressed as $1 -
2\frac{dist(f,\chi_a)}{2^n} = 1 - 2\frac{dist(f,\chi'_{a,0})}{2^n}$.
That is, $\frac{dist(f,\chi'_{a,0})}{2^n} = \half(1-\hat{f}(a))$.
Similarly, applying the same steps as above to the function $g(x) = f(x) \xor
1$, it can be shown that $-\hat{f}(a) = 1 -
2\frac{dist(f,\chi'_{a,1})}{2^n}$, i.e., $\frac{dist(f,\chi'_{a,1})}{2^n} =
\half(1 + \hat{f}(a))$.

Combining the expressions above leads us to another expression for $\nlf$:
$$\nlf = \min_a \Big\{ \thalf(1+\hat{f}(a)), \thalf(1-\hat{f}(a)) \Big\} = \min_{a} \thalf(1-|\hat{f}(a)|) = \thalf - \thalf \max_a |\hat{f}(a)| =
\thalf - \thalf \fmax.$$
This expression can be computed using the $\Theta(n2^n)$ fast-Walsh-Hadamard
transform algorithm and we get
a $\Theta(n2^n)$ exact algorithm for computing $\nlf$. However, this algorithm
has an additional overhead of $\Theta(2^n)$ space compared to the earlier
deterministic approach. To the best of our knowledge, there are no
better deterministic approaches known with asymptotically better time or space complexity for arbitrary Boolean functions. Therefore we focus on randomised, sampling-based, approaches. The subject of this paper is the {\tt non-linearity estimation} problem where we want to estimate with high probability the non-linearity with $\lambda$ additive accuracy, for any given $\lambda$.
\begin{equation}\label{eqn:goal}
    \boxed{\mbox{determine  $0 \le a < b \le 1$}\qquad \mathrm{such~that~} b-a \le \lambda \qquad\mathrm{and}\qquad
    \Pr[a < \nlf < b] \ge 1-\delta.}
\end{equation}

Recall that $\hat{f}(a)$ can also be expressed as the {\em expected} correlation of $f(x)$ and
$\chi_a(x)$ for $x$ sampled uniformly at random. \st{We can use standard approaches
for estimating the expected correlation; the number of samples of $Y$ (and
hence the running time complexity) to estimate $\E[Y]$} { Suppose we denote this correlation as $Corr(f)$. We can use standard approaches
for estimating $\E[Corr(f)]$; the number of samples of $x$ (and
hence the running time complexity) to estimate $\E[Corr(f)]$} with additive accuracy
$\lambda$ and error at most
$\delta$ will be $O(\frac{1}{\lambda^2} \log \frac{1}{\delta})$.
There is no additional space overhead in these approaches. However, since $\nlf$
depends upon the Fourier coefficient with the largest absolute value, the time and query complexity still runs into $\Theta(2^n)$. In this paper we design algorithms with query complexities that are polynomial in $(n,\lambda,\log \tfrac{1}{\delta})$.

Non-linearity is an important property of Boolean functions and trying to
compute or estimate it for a function, given either as a black-box or in some
other representation, is a natural question in the realm of Boolean functions.
However, beyond this academic curiosity lies the connection of non-linearity to
other hardness measures of Boolean functions. In a recent paper, Boyar et
al.~\cite{Boyar2016} considered 5 common measures apart from nonlinearity (algebraic
degree, annihilator immunity, algebraic thickness, normality, and multiplicative
complexity) and obtained relationships among them; for example, they show that
low multiplicative complexity implies low non-linearity and {vice versa}.
Many of {these} measures, including non-linearity, are used to design
cryptographic ciphers and hash-functions with interesting properties like
collision-resistance~\cite{Boyar2016} and propagation
characteristics~\cite{CCCF2000}. Even though we leave out these interesting applications
out of the scope of this paper, it would be worthwhile to understand the best use of a non-linearity estimation algorithm in
cryptography~\cite{sarkar2000nonlinearity}.

{ The design of our quantum algorithm could be of independent interest. We were recently able to use the idea therein to formulate quantum algorithms for a few variants of the element distinctness problem~\cite{our_highdist}.}

\subsection{Related work}

To highlight the computational challenge of computing, or even estimating, the
non-linearity of a Boolean function given as a black-box, recently
Bera et al.~\cite{BLRBera2019} investigated this question in the context of the
well-known BLR linearity testing algorithm~\cite{BLUM1993549}. The BLR test
evaluates a Boolean function given in the form of a black-box, always accepting
a linear function but sometimes accepting a non-linear function as well. They showed that
the probability of false-positive in a BLR test is not monotonic with
non-linearity, and hence, found it challenging to {\em compute} non-linearity by
employing BLR.

However, if we want to {\em estimate} non-linearity, allowing some inaccuracy,
then the above observation need not be a show-stopper.
Indeed, using $p$ to denote the probability that the BLR test accepts a function
$f$, it can be shown that $p = \thalf + \thalf \sum_a \hat{f}^3(a) \le \thalf +
\thalf \max_a \hat{f}(a)$.
Suppose one runs the BLR test multiple times to get a close estimate $\hat{p}$ of
$p$. Then it may be possible to estimate the lower bound 
$\max_a \hat{f}(a) \gtrapprox 2\hat{p}-1$. However, the trouble is that 
$2\hat{p}-1$ can be positive or negative and, if negative, we fail to get
any bound on $\fmax$.

Hillery et al.\ proposed {a property testing quantum algorithm} that makes $O(\tfrac{1}{\epsilon^{2/3}})$ queries~\cite{HilleryAnderssonPRA} and this was subsequently improved to $O(\tfrac{1}{\epsilon^{1/2}})$ queries~\cite{chakraborty2013improved}. 
But we faced hurdles when we tried to adapt these property testing algorithms { that identify} if a function $f$ is linear (i.e., $\nlf=0$) or is $\epsilon$-far from linear (i.e., $\nlf \ge \epsilon$). Since we don't consider promise problems {\em ala} property testing in this paper, so, if $f$ is neither linear nor $\epsilon$-far (for any guessed $\epsilon$), the algorithm may erroneously return ``linear'' or ``$\epsilon$-far'' and we get no insights whatsoever.

Consider the simpler problem of computing $x^*=\arg\max_x |\hat{f}^2(x)|$, and for simplicity, assume unique $x^*$; estimating $\nlf$ is easy with the knowledge of $x^*$ (a randomised algorithm for this is given by Lemma~\ref{thm:c_est_PWC} and a quantum algorithm is given by Lemma~\ref{thm:PWC}). It is known that the quantum circuit used in the Deutsch-Jozsa problem generates the state $\sum_x \hat{f}(x) \ket{x}$ which, when observed, gives us a state $\ket{x}$ sampled from the distribution $\{\Pr[x]=\hat{f}^2(x)\}$. Thus it is tempting to make multiple observations of independent runs of the Deutsch-Jozsa circuit and return the majority observation; the idea is that $x^*$ has the largest probability in the entire spectrum , and so, may have the largest probability among the observed samples. This is the scenario of using the mode of a few {\it i.i.d.} samples as an estimator of the mode of a discrete distribution. However, Dutta et al.\ showed that if $\min_{y \not= x^*} (\Pr[x^*]-\Pr[y]) \ge g(n)$, then the number of samples required is $O(\frac{n}{g^2} \log \frac{1}{\delta})$~\cite[Theorem 4]{modeestimation}. But this upper bound can be as large as exponential since we observed that $g(n)$ could be {$O(1/2^{1.5n})$} for $n$-bit functions~\footnote{Choose $f$ with Hamming distance 1 from a bent function, say $h$. It is straightforward to show that if $dist(f,h)=1$, then $\hat{f}(x) = \hat{h}(x) \pm \frac{2}{2^n}$; we obtain that the smallest gap between $\fmax^2$ and \st{$\{\hat{f}^2(x)\}$} { $\hat{f}^2_{\mathrm{2ndmax}}$} is \xcancel{$\frac{4}{2^{3n/2}}$} { $\frac{8}{2^{3n/2}}$}.}.

To the best of our knowledge, there exist very limited attempts
towards this problem, even considering the classical computing framework. We are aware of an algorithm for non-linearity computation of a sparse Boolean function (sparsity is with respect to the truth table)~\cite{calik2013}, however, neither that approach provides any 
accuracy guarantees nor it is designed in the usual black-box query model --- { there} the function is required to be given in its algebraic normal form.
In a recent pre-print~\cite{gowers_norm}, its authors related $\hat{f}_{max}^2$ to the Gower's $U_2$ norm of $f$ as $\lvert\lvert f \rvert\rvert^4_{U_2} \le \hat{f}^2_{max}$ and suggested that $\lvert\lvert f \rvert\rvert^4_{U_2}$ can be used to obtain a lower bound on $\hat{f}^2_{max}$ (which implies an upper-bound on non-linearity). {There is} a quantum circuit proposed by us in an earlier work~\cite{bera2019indocrypt} to estimate {$\lvert\lvert f \rvert\rvert^4_{U_2}$} with additive accuracy $\lambda$ using $\tilde{O}(\frac{1}{\lambda^2})$ queries~\footnote{The authors claimed a query complexity of 
$\tilde{O}(\frac{1}{\lambda})$~\cite[Equation~28]{gowers_norm}. However, they used an incorrect form of Hoeffding's inequality. Using the correct inequality gives a query complexity of $\tilde{O}(\frac{1}{\lambda^2})$ to obtain the estimate of the Gower's $U_2$ norm.}. However, this approach does not { control the accuracy of the estimate of } \st{even produce a lower bound on} $\fmax^2$ since there is no known theoretical \st{lower} { upper} bound on $\hat{f}^2_{max} - \lvert\lvert f \rvert\rvert^4_{U_2}$.

\subsection{Overview of results}

This paper resolves a few important questions in the light of the earlier discussions
 that non-linearity appears difficult without querying $f$ on exponentially-many inputs. The success of quantum query algorithms against Boolean functions motivated us to look into quantum algorithms. Can non-linearity be estimated with an additive constant inaccuracy using exponentially few queries to $f$? Or, even constant many queries? What is the minimum number of queries needed if accuracy is not a constant? What about classical randomised algorithms? After all, the BLR test is pretty effective.
 
 Our techniques are primarily quantum in nature, but we also obtain results for randomised algorithms along the way. Here we are interested in query complexity, and so our algorithms require access
 to a unitary representation of a Boolean function $f$ (denoted \st{by} $U_f$).
$\lambda$ denotes the accuracy parameter and $\delta$ denotes the maximum
allowed probability of error. 
Our results are summarised in Table~\ref{table:results}.


\begin{table}[!h]
    \centering
    \caption{Our results on estimating $\nlf$ with additive error $\lambda$ (ignoring logarithmic factors). {\em Note the independence of $n$ in the quantum algorithm complexity.}~\label{table:results}}
\begin{tabular}{|l||c|c|c|}
    \hline
     Worst-case & \multicolumn{2}{c|}{Algorithm} & Lower bound \\
     complexity & Query complexity & Number of qubits & Query complexity\\
     \hline
     Randomised & & & \\
     algorithm & $O(n/\lambda^6)$ [Theorem~\ref{thm:cnonlinalgo}] & - & $\Omega(1/\lambda)$ [Theorem~\ref{thm:rlb}]\\
    \hline
     Quantum & & & \\
     algorithm & $O(1/\lambda^3)$ [Theorem~\ref{thm:nonlinalgo}] & $O\Big(\log{\big(\frac{1}{\delta\lambda}\big)}\cdot \big(n+\log\big(\frac{1}{\lambda^2}\big)\big) \Big)$ & $\Omega(1/\sqrt{\lambda})$ [Theorem~\ref{cor:nonlin_lower}]\\
     \hline
\end{tabular}
\end{table}

Our algorithms are most suitable for estimating non-linearity up to a constant or poly-logarithmic bits of precision (the smallest non-zero non-linearity requires roughly $n/2$ bits after the decimal point). 
It is easy to show that $|\hat{f}(x)-\hat{f}(y)| \ge \frac{2}{2^n}$ whenever $\hat{f}(x)$ and $\hat{f}(y)$ are distinct. Therefore, non-linearity can be exactly computed if we set $\lambda=2/2^n$. Our lower bounds say that to compute non-linearity exactly, classically there is nothing better than querying $f$ at all the $2^n$ points; however, a $O(2^{n/2})$-query quantum algorithm probably exists.


A notable feature of our algorithm is that, unlike the classical ``fast Walsh-Hadamard transformation'' approaches, our algorithms are iterative in nature requiring little additional space. Further, there is very little overhead in the running time on top of the queries to $U_f$. Hence the query complexity above directly translates to its time complexity as well, { with $poly(n)$ overhead arising from the additional gates required to perform amplitude estimation, amplification and small sub-circuits.}


\subsection{Overview of techniques}

We repeatedly estimate the probability $p$ of an observation upon measuring the final state of a quantum circuit $A$. Using quantum amplitude estimation~\cite{brassard2002quantum} we can obtain 
an estimate $\tilde{p}$ such that
$\Pr[|p-\tilde{p}| \ge \epsilon] \le \delta$ for any $\epsilon \le
\frac{1}{4}$ using a total of $\Theta(\frac{1}{\epsilon}\log\frac{1}{\delta})$
calls to $A$ (details given as Corollary~\ref{cor:ampest} in Appendix~\ref{appendix:ampest}). Note that a classical algorithm for the same task would require $O(\frac{1}{\epsilon^2}\log\frac{1}{\delta})$ calls to $A$, or as we would often state, $\tilde{O}(\tfrac{1}{\epsilon^2})$ calls after ignoring $\log$-factors.

 We also use quantum amplitude amplification, in particular, its fixed point version~\cite{fixedpointaa} which gives us a quantum circuit that when measured at the end gives us a good state, if any such exists. If the probability of the original algorithm is lower bounded by $p$, then the number of iterations, hence queries, can be upper bounded by $O(\sqrt{1/p})$. The fixed point version could have been replaced with other amplitude amplification variations that require only a lower bound on the success probability~\cite{brassard2002quantum}; however, we \st{require} { prefer} the fixed point version since it does not involve any intermediate measurement { and can be used inside another amplitude amplification.}


Both our randomised and quantum algorithms to estimate non-linearity with inaccuracy $\lambda$ actually estimate $\fmax^2$ with inaccuracy $\Theta(\lambda^2)$. The latter is implemented as a binary search, named {\tt IntervalSearch}, to find the {\em largest threshold} $\tau$, among a discrete set of thresholds that depend on $\lambda$, such that $\fmax^2 \ge \tau$. The difficulty lies in solving the \bfmax decision problem that, given $\tau$, decides if $\fmax^2 \ge \tau$ in sub-exponential time. There is a technical challenge in getting binary search to act since the algorithms use estimations to guide the search. The estimations have an additive error, and we have to be careful during the comparisons with the estimated values made by the binary search. Our technical contributions here are a randomised algorithm and a quantum algorithm for the \bfmax problem.

The classical randomised approach, named {\tt CBoundFMax}, searches among all $\hat{f}(x)^2$ values; however, it uses the idea of the Goldreich-Levin algorithm~\cite{goldreich1989hard} to restrict search among only a small subset of values. Value of any specific $\hat{f}(x)^2$ is of course not readily available, but that can be easily estimated using a few $f(x)$ values sampled randomly. The number of queries is linear in $n$ and scales inversely with $\tau^3$. It is possible to convert this algorithm to a quantum one, but we would end up with a complexity that scales as $\tfrac{n}{\tau^2}$ --- indeed, that is the complexity of the quantum versions of the Goldreich-Levin algorithms that have been proposed so far~\cite{li2019quantum_gl,montanaro2010quantum}.

To understand how {\tt QBoundFMax} gets rid of the dependence on $n$, it will be useful to understand {\tt CBoundFMax}. Think of $\epsilon$ to be something that is smaller than $\tau$, say, $\tau/2$. {\tt CBoundFMax} performs a level-order traversal of a binary tree built on all possible binary prefixes of length up to $n$. At any particular node, say $p$, the algorithm estimates $PWC(p^\frown 0)$ where $PWC$ at any prefix $q$ is defined as $PWC(q)=\sum_{x\in\{0,1\}^{n-|q|}} \hat{f}^2(q^\frown x)$ { where $q^\frown x$ denotes the concatenation of $q$ with $x$}. Note that for $PWC(p^\frown 0)$, the summation is over all $n$-bit $x$ with prefix $p^\frown 0$\st{ where $p^\frown 0$ denotes the concatenation of $p$ with $0$} and $PWC(p^\frown 1)$ is defined similarly. It identifies those 1-bit extensions of $p$ for which {$PWC(p^\frown b)\ge \tau - \epsilon$} and adds them to a queue. Once all the nodes of a level is processed, the nodes in the queue are retrieved and processed in the manner described above.
At the final level $l=n$, $PWC(p)=\hat{f}^2(p)$ for any $n$-bit prefix $p$. If any prefix at the final level satisfy $PWC(p)\ge \tau$, then the algorithm concludes that $\hat{f}^2_{max} \ge \tau$.

It is immediate that the {\tt CBoundFMax} algorithm has three components that contribute to its complexity; a) the estimation of $PWC(p)$ with additive accuracy $\epsilon$ that takes $O(\frac{1}{\epsilon^2})$ queries, b) the total number of prefixes added to the queue at any particular level $l$ which is {$O(\frac{1}{\tau-2\epsilon})$} by applying Parseval's identity, and c) the outer loop for the level-order traversal which is $O(n)$. 
Hence the total query complexity of {\tt CBoundFMax} is {$O(\frac{n}{\epsilon^2(\tau-2\epsilon)})$ queries}.

The {\tt CBoundFMax} algorithm can be improved if we reduce the complexity of any of the three components of {\tt CBoundFMax} and replace the classical loops and data structures with their quantum equivalent ones.
The most trivial way to improve the complexity is to replace the classical estimation with the quantum estimation. While classical estimation uses $O(\frac{1}{\epsilon^2})$ queries, its quantum counterpart (implemented using the Deutsch-Jozsa circuit) uses $O(\frac{1}{\epsilon})$ queries leading to the final query complexity {$O(\frac{n}{\epsilon(\tau-2\epsilon)})$} (this algorithm is explained in Appendix~\ref{subsec:cboundfmax-qpwc}). This is exactly what has been proposed earlier as the quantum version of Goldreich-Levin~\cite{montanaro2010quantum,li2019quantum_gl} in which we set $\tau=\lambda^2$ and $\epsilon=\tau/2$ to get a list of all $x$ such that $\fmax^2(x) \ge \lambda^2$.

The above algorithm runs a classical subroutine around a quantum circuit (for estimating $PWC(p)$ on an eligible $p$) in each level. Its query complexity can be improved by using a single quantum circuit for the entire operations of a level: (i) Estimation of $PWC()$, followed by (ii) filtering based on comparison with $\tau$. This can be implemented by generating a superposition of all eligible $p$ in a level, say $\sum_p c(p) \ket{p}$ where $|c(p)|^2 \approx PWC(p)$, running amplitude estimation without the measurement to store $|c(p)|^2$ in some register and then comparing the value in this register to that of $\tau$ to mark some of the $\ket{p}$s in the superposition (see Appendix~\ref{subsec:qcboundfmax} for the entire algorithm). The dependence of the query complexity on $n$ remains there, but it nevertheless improves to {$\tilde{O}\left(\frac{n}{\epsilon\sqrt{\tau-2\epsilon}}\right)$.}


To remove the dependence on $n$, we remove the classical level-order traversal altogether and replace the equally superposed initial state by an initial state in which each basis state has amplitude proportional to its Walsh coefficient. Using a clever combination of Deutsch-Jozsa, amplitude amplification and amplitude estimation, {\tt QBoundFMax} manages to achieve a complexity of {$\tilde{O}\left(\frac{1}{\epsilon\sqrt{\tau-2\epsilon}}\right)$} queries. What is remarkable is that the final algorithm can be implemented as a single quantum circuit (see Figure~\ref{fig:qbound_fmax}) unlike many quantum algorithms which are essentially classical wrappers around amplitude amplification and amplitude estimation.

\section{Interval search for $\hat{f}^2_{max}$}\label{sec:interval_search}

In this section, we consider the problem of estimating an interval $J \subseteq (0,1)$
of length $|J| \le \epsilon$ such that $\hat{f}_{max}^2 \in J$ with high probability. We will use this as stepping stone for estimating $\hat{f}_{max}^2$ with any desired additive accuracy $\epsilon$. Let $k$ be the smallest integer such that $\frac{1}{2^k} \le \epsilon/2$. For finding $J$, our {\tt IntervalSearch} algorithm for the above problem divides the
interval $[0,1]$ into sub-intervals of length \xcancel{$\frac{1}{2^{k+1}}$} {$\frac{1}{2^{k}}$} and then finds the
right-most (i.e., towards 1) sub-interval that contains {\em any} non-zero Fourier
coefficient-squared (i.e., $\hat{f}^2(a)$ for any $a$). Clearly, $\hat{f}_{max}^2$ must belong to the same interval.

To implement the above strategy, we need to first solve the following problem that we call as \bfmax: Given a function $f(x)$ as a blackbox, a threshold $\tau \in (0,1)$, and accuracy $g$ (we will refer to this as the ``gap'') perform the following with probability of error at most $\delta$.
\begin{itemize}
    \item Return {\tt TRUE} if $\hat{f}^2_{max} \ge \tau$.
    \item Return {\tt FALSE} if $\hat{f}^2_{max} < \tau-2g$.
    \item Return anything if $\hat{f}^2_{max} \in [\tau-2g, \tau)$.
\end{itemize}

Classical and quantum algorithms for the \bfmax problem are explained in Sections~\ref{sec:classical_algo} and \ref{sec:qbound_fmax}, respectively. For now, assume that we have an homonymous algorithm for the problem.

The {\tt IntervalSearch} algorithm is described in Algorithm~\ref{algo:intervalsearch}. It uses {\em binary
search} to find the rightmost \st{(i.e., with highest boundary values)} interval { (i.e., with highest boundary values)}
among all the $2^k$ sub-intervals that contain $\fmax^2$. It uses {\tt
BoundFMax} to decide whether to consider the
``right-half'' or ``left-half'' of the currently processing interval.
To handle the third case of {\tt BoundFMax}, we equipped {\tt IntervalSearch} to search among slightly overlapping intervals.

    \begin{algorithm}
	\caption{Algorithm {\tt IntervalSearch} to find out an $\epsilon$-length
	interval containing $\fmax^2$\label{algo:intervalsearch}}
	\begin{algorithmic}
	    \Require accuracy $\epsilon$ and probability of error $\delta$
	    \State Set $k = \left\lceil \log_2 \tfrac{1}{\epsilon}
	    \right\rceil + 1$ \Comment{$k$ is the smallest integer {\it s.t.}
	    $\tfrac{1}{2^k} \le \frac{\epsilon}{2}$; thus, $\frac{\epsilon}{4}
	    < \frac{1}{2^k} \le \frac{\epsilon}{2}$}
	    \State Set gap $g=\frac{1}{8}\left(\epsilon - \tfrac{1}{2^k}\right)$
	    \Comment{$8g + \frac{1}{2^k} = \epsilon \implies \tfrac{3}{2}\tfrac{\epsilon}{16} \ge g \ge \tfrac{\epsilon}{16}$}
	    \State Set boundaries $lower=\tfrac{1}{2^n}$, $upper=1$ and threshold
	    $\tau=\frac{1}{2}$
	    \For{$i=1 \ldots k$}
		\If{{\tt BoundFMax}$(\tau, g, \frac{\delta}{k}) \to \mathtt{TRUE}$}
		    \State Update $lower = \tau - 2g$, $\tau = \tau +
	    \tfrac{1}{2^{i+1}}$; $upper$ is unchanged
		\Else
		    \State Update $upper = \tau$, $\tau = \tau -
	    \tfrac{1}{2^{i+1}}$; $lower$ is unchanged
		\EndIf
	    \EndFor
	    \State \Return $[lower,upper)$
	\end{algorithmic}
    \end{algorithm}

\begin{figure}[!ht]
    \resizebox{\linewidth}{!}{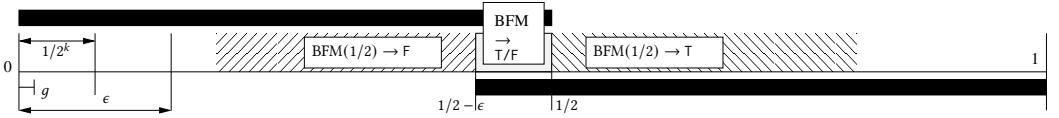}
    \caption{Explanation of the first binary-search step of the {\tt IntervalSearch} algorithm for a scenario when $\epsilon$ is a power of 2, say $2^{2-k}$. BFM$(1/2)$ denotes the output of the {\tt
    BoundFMax} algorithm using $\tau=\frac{1}{2}$. The black rectangles indicate
    the two possible ranges that will be searched next depending upon {\tt
    BoundFMax}$(1/2)$. { The shaded regions indicate the outputs of the BFM calls for different values of $\fmax^2$.} \label{fig:intervalsearch}}
\end{figure}

It will be easier to understand {\tt IntervalSearch} from the
illustration given in Figure~\ref{fig:intervalsearch} in which we have assumed
$\epsilon$ to be a power of $\tfrac{1}{2}$; in this case $\frac{1}{2^k}=\tfrac{\epsilon}{2}$
and the gap $g$ used for {\tt BoundFMax} is $\tfrac{\epsilon}{16}$.
It is easy to verify that the search interval $[lower,upper)$ before round 1 has
length $1-\frac{1}{2^n}$ and before round $i$ has length $\frac{1}{2^{i-1}} - {\tfrac{1}{2^n}}$ (if {\tt BoundFMax}
has returned {\tt FALSE} \st{is} { in} all the previous $i-1$ rounds) or $\frac{1}{2^{i-1}}
+ 2g$ (if {\tt BoundFMax} has returned {\tt TRUE} in any of the previous rounds).
Note that $\tau$ is always `midway' of a search interval excluding $g$, the overhead due to the gap.

If {\tt BoundFMax}
returns {\tt TRUE}, then we are sure that $\fmax^2$ lies in the right-half of
the interval (with a slight overhead of $2g$ at the lower boundary) --- accordingly, the
lower-boundary and the threshold are moved right. On the other hand if {\tt
BoundFMax} returns {\tt FALSE}, then we are sure that $\fmax^2$ lies in the
left-half of the interval; so, the upper-boundary and the threshold are moved
left. Thus, before and after each round it is ensured that $lower \le \fmax^2 <
upper$ for the current values of $lower$ and $upper$.

Binary search ends after the $k$-th round. The interval contains $\fmax^2$ and
its length is at most $\frac{1}{2^k}+2g=\epsilon - 6g < \epsilon$.

\begin{theorem}\label{thm:int_search}
Let $f$ be an $n$-bit Boolean function. Given an additive accuracy $\epsilon$ and probability of error $\delta$, there is a quantum algorithm of query complexity $\tilde{O}\left(\frac{1}{\epsilon^{3/2}} \right)$ and a classical algorithm of query complexity $\tilde{O}\left( \frac{n}{\epsilon^{3}} \right)$ that 
outputs an interval of length at most $\epsilon$ that, with probability at
least $1-\delta$, contains $\fmax^2$.
\end{theorem}

\begin{proof}
The Algorithm {\tt IntervalSearch} makes $k$ invocations to {\tt BoundFMax}. Let $c_b(\tau,g,\delta{/k})$ be the number of calls that {\tt BoundFMax} makes to the oracle to solve its problem with threshold $\tau$, accuracy $g$ and error  $\delta{/k}$. 
Then, the total number of calls to $U_f$ is upper bounded
by $\sum_{i=1}^k c_b(\tau_i, g,\delta{/k})$ in which $\tau_i$
denotes the value of $\tau$ in the $i$-step of the binary-search. 

Using the classical implementation of \bfmax (Lemma~\ref{thm:cboundfmax}), we have $\sum_{i=1}^k c_b(\tau_i, g,\delta) = \frac{n}{g^2}
\Ot\Big(\sum_{i=1}^k \frac{1}{\tau_i - 2g}\Big)$.
The way {\tt
IntervalSearch} sets $k$ and $g$ ensures that $\frac{\epsilon}{2} \le 8g = \epsilon - \frac{1}{2^k} <
\frac{3\epsilon}{4}$. Furthermore, all $\tau_i \ge \frac{1}{2^k} >
\frac{\epsilon}{4} \implies \tau_i - 2g > \frac{\epsilon}{16}$.
    This leads to the classical query complexity of $\Ot \left( \frac{n}{(\epsilon/16)^2 \cdot \epsilon/16} \right) = \Ot(\frac{n}{\epsilon^3})$ { further} hiding a $O(\log\tfrac{1}{\epsilon})$ factor.

With the quantum implementation of {\tt BoundFMax} (Lemma~\ref{lemma:QBoundFMax}) we have  $\sum_{i=1}^k c_b(\tau_i, g,\delta) =\\ \frac{1}{g}
\Ot\Big(\sum_{i=1}^k \frac{1}{\sqrt{\tau_i - 2g}}\Big)$.
Therefore,
we get the number of calls as $\Ot \left( \frac{1}{(\epsilon/16) \cdot
\sqrt{\epsilon/16}} \right) = \Ot(\frac{1}{\epsilon^{3/2}})$.

As for the error, there are $k$ calls to {\tt BoundFMax} that is allowed to return an incorrect
answer with probability at most $\frac{\delta}{k}$. Therefore, there is an
overall probability of $\delta$ that any of those calls return an incorrect
answer.
\end{proof}

\section{Classical randomised algorithm for \bfmax}\label{sec:classical_algo}

Goldreich and Levin proposed a randomised algorithm for learning the ``high'' Walsh coefficients of a Boolean function~\cite{goldreich1989hard}. Our randomised algorithm follows the presentation of this algorithm as a level-order traversal of a binary tree. We borrow from the book by O'Donnell~\cite{odonnell-book} the definition $\mathbf{W}^{S|\bar{J}}[f]$, denoted here as PrefixWalshCoefficients (in short, PWC) 
of $f$ at a prefix $a$, and Proposition~3.40 as the forthcoming lemma .

\begin{definition}[PrefixWalshCoefficients]
    For a prefix $a \in \{0,1\}^s$ such that $0\le s \le n$, $PWC(a)=\sum_{x \in \{0,1\}^{n}} \hat{f}^2(x)$ where the summation is over all $n$-bit $x$ with prefix $a$.
\end{definition}

It immediately follows that if $a$ is $n$-bit, then $PWC(a)=\hat{f}^2(a)$.

\begin{lemma}[\cite{odonnell-book}]
    \label{thm:c_est_PWC}
    There is a $O\left(\frac{1}{\epsilon^2}\log \frac{1}{\delta}\right)$-query classical algorithm, { denoted $PWCE$,} for estimating $PWC(a)$ for a prefix $a$ within $\pm\epsilon$ and with probability at least $1-\delta$. In particular, this algorithm can estimate $\hat{f}^2(b)$ for an $n$-bit $b$.
\end{lemma}

We include a proof of the lemma for completeness.
\begin{proof}

$\hat{f}(b)$ can be expressed as $\E_{x} (-1)^{f(x) \oplus x \cdot b}$; so it can be estimated by simply averaging $(-1)^{f(x) \oplus x \cdot b}$ for some uniformly chosen random $x \in \{0,1\}^n$ --- the number of queries required follow from Hoeffding's bound for $\pm 1$ random variables.
    
    Estimating $PWC(a)$ requires expressing it too as the expectation of a $\pm 1$ random variable as shown below; here the length of $a$ is denoted $k$.

    \begin{align*}
	PWC(a) & = \sum_{b\in\{0,1\}^{n-k}} \hat{f}(ab)^2 = \tfrac{1}{2^{2n}}\sum_{b \in \{0,1\}^{n-k}} \bigg( \sum_{x \in \{0,1\}^n} (-1)^{f(x) \oplus x\cdot (ab)} \bigg)^2\\
	& = \tfrac{1}{2^{2n}}\sum_{b \in \{0,1\}^{n-k}} \bigg( \sum_{x,y \in \{0,1\}^n} (-1)^{f(x) \xor f(y) \oplus x\cdot (ab) \xor y \cdot (ab)} \bigg)\\
	& = \tfrac{1}{2^{2n}} \sum_{\substack{x_1,y_1 \in \{0,1\}^k\\x_2, y_2 \in \{0,1\}^{n-k}}} \sum_{b \in \{0,1\}^{n-k}} (-1)^{f(x_1x_2) \xor f(y_1y_2) \xor (x_1x_2)\cdot(ab) \xor (y_1y_2)\cdot (ab)}\tag{Split $x=x_1x_2$, $y=y_1y_2$}\\
	& = \tfrac{1}{2^{2n}} \sum_{\substack{x_1,y_1 \in \{0,1\}^k\\x_2, y_2 \in \{0,1\}^{n-k}}}  (-1)^{f(x_1x_2) \xor f(y_1y_2) \xor x_1\cdot a \xor y_1 \cdot a} \sum_{b \in \{0,1\}^{n-k}} (-1)^{(x_2 \xor y_2) \cdot b} \\
	& = \tfrac{2^{n-k}}{2^{2n}} \sum_{\substack{x_1,y_1 \in \{0,1\}^k\\x_2 = y_2 \in \{0,1\}^{n-k}}}  (-1)^{f(x_1x_2) \xor f(y_1x_2) \xor x_1\cdot a \xor y_1 \cdot a} \tag{$\sum_b (-1)^{z \cdot b}$ is 0 if $z\neq 0$, else $2^{n-k}$}\\
	& = \tfrac{1}{2^{n-k} 2^{k} 2^k} \sum_{\substack{x_1,y_1 \in \{0,1\}^k\\x_2 \in \{0,1\}^{n-k}}}  (-1)^{f(x_1x_2) \xor f(y_1x_2) \xor x_1\cdot a \xor y_1 \cdot a} \\
	& = \E_{x_1,y_1,x_2} \Big[ (-1)^{f(x_1x_2) \xor f(y_1x_2) \xor x_1\cdot a \xor y_1 \cdot a} \Big]
    \end{align*}
\end{proof}


Our classical algorithm {\tt CBoundFMax} for the \bfmax problem is sketched in Algorithm~\ref{algo:cbound_fmax}.

    \begin{algorithm}[!h]
	\caption{\label{algo:cbound_fmax}Algorithm \texttt{CBoundFMax}}
	\begin{algorithmic}
	    \Require threshold $\tau \in (0,1)$, confidence
	    $\epsilon \in (0,\tau)$, error $\delta \in (0,1)$
	    \State Initialise a FIFO list $Q = \{\varepsilon\}$, where $\varepsilon$ denotes the empty string
	    \While{$Q$ is not empty}
		\State remove prefix $\mathfrak{p}$ from $Q$
		\For{suffix $\mathfrak{s}$ from $\{$ 0, 1$\}$}
		    \State childprefix $\fcp = \mathfrak{p^\frown s}$
		    \State obtain estimate $e \leftarrow
	    \pwce(\fcp, \epsilon, \left( \frac{\tau-\epsilon}{2n}\right) \delta)$ \Comment{With accuracy $\epsilon$ and error $\delta'=\frac{\tau-\epsilon}{2n} \delta$}
		    \If{${e} \ge \tau-\epsilon$}
			\State If $len(\fcp) < n$, add $\mathfrak{cp}$
			to $Q$
			\State Else (i.e., $len(\fcp) = n$), \Return
	    {\tt TRUE}
		    \EndIf
		\EndFor
	    \EndWhile
	    \State \Return {\tt FALSE}
	\end{algorithmic}
    \end{algorithm}

\begin{lemma}\label{thm:cboundfmax}
    Algorithm {\tt CBoundFMax} solves the \bfmax problem using $\tilde{O}(\frac{n}{\epsilon^2 (\tau-2\epsilon)})$ queries.
\end{lemma}

\begin{proof}
The algorithm traverses in level-order a binary tree on all strings of lengths up to $n$ to find some $x$ such that $\hat{f}^2(x) \ge \tau$; children of a node $a \in \{0,1\}^s$ are denoted $a^\frown 0$ and $a^\frown 1$.

It uses the observation that if $PWC(a) = \sum_{b \in \{0,1\}^{n-s}} \hat{f}^2(a^\frown b)$ is less than $\tau - \epsilon$, then there cannot be any $x$ with prefix $a$ for which $\hat{f}^2(x) \ge \tau - \epsilon$ and hence the subtree under $a$ need not be further explored. However there may be an inaccuracy in estimating $PWC(a)$ which we handle in the two cases below.

\paragraph*{Case --- {\tt CBoundFMax} returns {\tt TRUE}:} This happens only when
the algorithm finds some $n$-bit $\fcp$ for which the $\pwc$ estimate $e$
satisfies $e \ge \tau - \epsilon$. From Lemma~\ref{thm:c_est_PWC}, $e$
satisfies $\pwc(\fcp)-\epsilon \le e \le \pwc(\fcp) + \epsilon$
with probability at least $1-\delta'$. Furthermore,
$\pwc(\fcp)=\hat{f}^2(\fcp)$. Combining all these results, we see
that the following holds with high probability.
$$\tau \le e + \epsilon \le (\pwc(\fcp) + \epsilon) + \epsilon =
\hat{f}^2(\fcp) + 2\epsilon \qquad \implies \qquad \fmax^2 \ge \tau -
2\epsilon$$

\paragraph*{Case --- {\tt CBoundFMax} returns {\tt FALSE}:} For this
case, assume that on the contrary $\fmax^2 \ge \tau$, i.e., there is
some $n$-bit $x$ for which $\hat{f}^2(x) \ge \tau$. Therefore, for
all prefixes $\fp$ of $x$, $\pwc(\fp) \ge \hat{f}^2(x) \ge \tau$.
From Lemma~\ref{thm:c_est_PWC}, with probability at least $1-\delta'$, $e$
satisfied $\pwc(\fp)-\epsilon \le e \le \pwc(\fp) +
\epsilon$. For a moment assume that the estimator of $\pwc$ makes no error; thus, $e \ge \tau - \epsilon$
and when that holds, $\fp$ is added to $Q$ and eventually retrieved and
processed. Since the above fact holds for all prefixes of $x$, so, all
of them will be stored in $Q$ and retrieved which means that $x$ will
also be added to $Q$ and retrieved and processed. When $\fp=x$, $e \ge
\pwc(x) - \epsilon =\hat{f}^2(x) - \epsilon \ge \tau - \epsilon$.
Thus, {\tt CBoundFMax} should be returning {\tt TRUE} when that happens
--- this leads to a contradiction. Therefore, when {\tt CBoundFMax}
returns {\tt FALSE} it must be true that, with high probability,
$\fmax^2 < \tau$.

Let $\pwce$ be the classical estimator for $\pwc$.
The total number of calls to $\pwce$ will be at most $2 \times
\left( \frac{n}{\tau - 2\epsilon} \right)$ --- there are two suffixes to
try for each prefix and due to Parseval's identity, for every length $s \in \{1, \ldots n\}$, there are at most $\frac{1}{\tau - 2\epsilon}$ prefixes $\{\fp_1, \fp_2, \ldots \}$ of length $s$ such that $PWCE(\fp_i,\epsilon,\delta') \ge \tau - \epsilon$ where $\delta'$ is some error.
This is because of the fact that for any prefix $\fp$, $\pwc(\fp) < \tau-2\epsilon \implies \pwce(\fp,\epsilon,\delta) < \tau-\epsilon$. So we have $\pwce(\fp,\epsilon,\delta')\ge \tau-\epsilon \implies \pwc \ge \tau-2\epsilon$. 
Since at any level the total number of prefixes $\fp$ such that $\pwc(\fp)\ge \tau-2\epsilon$ is at most $\frac{1}{\tau-2\epsilon}$, we have that the total number of prefixes $\fp$ such that $\pwce(\fp,\epsilon,\delta')\ge \tau-\epsilon$ is at most $\frac{1}{\tau-2\epsilon}$.
The query-complexity is obtained by combining the number of calls to $\pwce$ with Lemma~\ref{thm:c_est_PWC}.

The algorithm works in a flawless manner as described if all the $\pwce$ calls are within their promised accuracy with no error. Since $\pwce$ is called with error parameter $\delta'=\left(
\frac{\tau-2\epsilon}{2n}\right) \delta$ therefore the probability of
{\tt CBoundFMax} facing any error is at most
$\delta$.
\end{proof}

\section{Quantum algorithm for \bfmax}\label{sec:qbound_fmax}

In the previous section we obtain a classical algorithm to solve the \bfmax problem using $O(\frac{n}{\epsilon^2(\tau - 2\epsilon)})$ queries. Note that the estimation of \pwc in Algorithm~\ref{algo:cbound_fmax}  is done classically.
A naive approach \st{one could use} to reduce the complexity of the algorithm is \st{by replacing} {to replace} the classical estimation by a quantum algorithm for \pwc estimation.

This quantum algorithm simply executes the
Deutsch-Jozsa circuit and measure the first $s$ qubits in the standard basis. The
probability of observing $\ket{a}$ is $\sum_{b \in \{0,1\}^{n-s}}
\hat{f}^2(ab)= PWC(a)$.
It directly follows that $PWC(a)$ can be estimated using amplitude estimation. The number of calls to $U_f$ that is required to achieve additive accuracy $\epsilon$ and probability of error $\delta$ is $O\left(
\frac{1}{\epsilon} \log \frac{1}{\delta} \right)$.

\begin{lemma}\label{thm:PWC}
    There is a $O\left(\frac{1}{\epsilon}\log \frac{1}{\delta}\right)$-query quantum algorithm for estimating $PWC(a)$ for a prefix $a$ within $\pm\epsilon$ and with probability at least $1-\delta$. The algorithm computes $\hat{f}^2(b)$ for an $n$-bit $b$ .
\end{lemma}

Using the quantum estimation subroutine for $PWC$ estimation inside Algorithm~\ref{algo:cbound_fmax} gives us a simple quantum algorithm for the \bfmax problem. The number of queries can easily be shown to be $\tilde{O}(\frac{n}{\epsilon(\tau - 2\epsilon)})$ which is already better compared to the classical algorithm discussed in Lemma~\ref{thm:cboundfmax}.
We now explain how to remove the dependency on $n$ and improve the dependency on $(\tau - 2\epsilon)$.

\begin{figure}
    \centering
    \includegraphics[width=\linewidth]{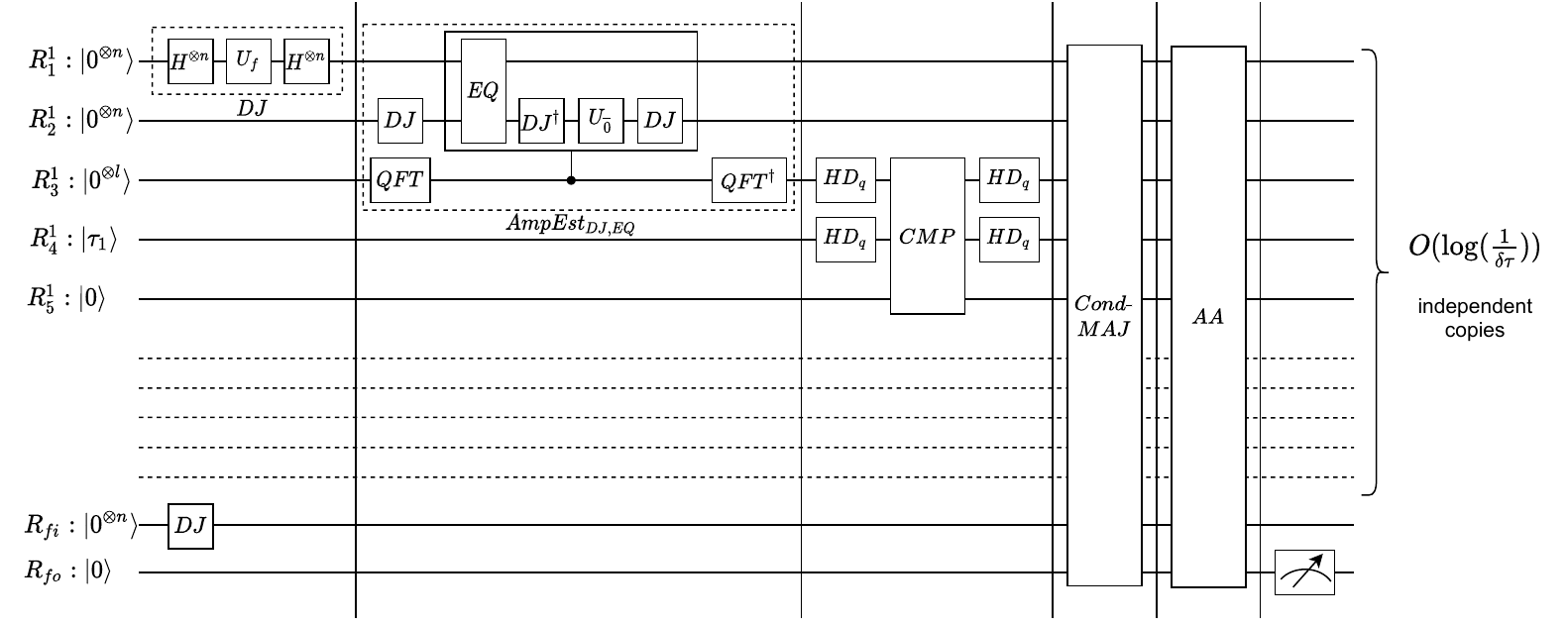}
    \caption{The {\tt QBoundFMax} circuit. The four stages are separated using vertical lines.}
    \label{fig:qbound_fmax}
\end{figure}

Our quantum algorithm {\tt QBoundFMax} is described in  Algorithm~\ref{algo:qbound_fmax} and a quantum circuit for the same is illustrated in Figure~\ref{fig:qbound_fmax}. 
It is to be noted that $R_i$ is used to represent the $i^{th}$ register used in the quantum circuit corresponding to the algorithm.
We use $DJ$ to represent the circuit for Deutsch-Jozsa: $H^{\otimes n}\cdot U_f \cdot H^{\otimes n}$ and we use a few smaller circuits listed below.

\begin{algorithm}
    \caption{Algorithm \texttt{QBoundFMax} \label{algo:qbound_fmax}}
    \begin{algorithmic}[1]
        \Require Threshold $\tau$, accuracy $\epsilon$ and error $\delta$.
        \State Set $\tau' = \tau - \frac{\epsilon}{8}$, $q = \lceil \log(\frac{1}{\epsilon}) \rceil +4$ and $l= q+3$.
        \State Set $\tau_1 = \left\lfloor{\frac{2^l}{\pi}\sin^{-1}(\sqrt{\tau'})}\right\rfloor$
        \On{$21\ln(\frac{1}{\delta^2\tau^2})$ many independent copies} \Comment{index $i$ ranges from 1 to $21\ln(\frac{1}{\delta^2\tau^2})$}
            \State Initialize $5$ registers $R^i_1R^i_2R^i_3R^i_4R^i_5$ as $\ket{0^n}\ket{0^n}\ket{0^{l}}\ket{\tau_1}\ket{0}$.
            The $4^{th}$ register is on $l$ qubits.
            \State {\bf Stage 1:} Apply $DJ = H^{\otimes n} \cdot U_f \cdot H^{\otimes n}$ on $R^i_1$.
            \State {\bf Stage 2:} Apply quantum amplitude estimation sans measurement ($AmpEst$) on
            $DJ$ with $R^i_2$ as the input register, $R^i_3$ as the precision register and 
            $R^i_1$ is used to determine the ``good state''. $AmpEst$ is called with 
            error at most $1- \frac{8}{\pi^2}$ and additive accuracy $\frac{1}{2^q}$.\label{line:amp_est}
            \State {\bf Stage 3:} Use ${\tt HD_l}$ on $R^i_3$ and $R^i_4$
            individually.\label{line:half_dist}
            \State Use ${\tt CMP}$ on $R^i_3$ and $R^i_4 = \ket{\tau_1}$ as input registers 
            and $R^i_5$ as output register.\label{line:q_compare} 
            \State Use ${\tt HD^{\dagger}_l}$ on $R^i_3$ and $R^i_4$
            individually again.
        \EndOn
        \State {\bf Stage 4:} Initialize two new registers $R_{fi}R_{fo}$ as $\sum_{x\in\{0,1\}^n}\hatf{x}\ket{x}\ket{0}$.
        \State For each basis state $\ket{x}$ in $R_{fi}$, for $i=1 \ldots 21\ln(\frac{1}{\delta^2\tau^2})$ compute the majority of the basis states of each $R^i_5$ register conditioned on the corresponding $R^i_1$ to be in $\ket{x}$, and store the result in $R_{fo}$.
        \State {\bf Stage 5:} Apply Fixed Point Amplitude Amplification 
        (AA) $\frac{1}{\sqrt{\tau-2\epsilon}}\log(\frac{2}{\sqrt{\delta}})$ times on $R_{fo}$ and measure $R_{fo}$ as $m$.
        \If{$m = \ket{1}$}
            \State \Return {\tt TRUE}
        \Else
            \State \Return {\tt FALSE}
        \EndIf \label{line:finish}
    \end{algorithmic}
\end{algorithm}

\begin{description}
    \item[{\tt EQ}:] Checking for equality of two $n$-bit strings, $EQ$ maps $2n$-qubit basis states $\ket{x,y}$ to $(-1)\ket{x,y}$ if $x=y$ and does nothing otherwise.
    \item[{\tt HD$_q$}:] When the target qubit is $\ket{0^q}$, and with a $q-$bit string $y$ in the control register, {\tt HD} computes the absolute difference of $y_{int}$ from $2^{q-1}$ and outputs it as a string where $y_{int}$ is the integer corresponding to the string $y$. It can be represented as ${\tt HD}_q \ket{y}\ket{b} = \ket{b\oplus\tilde{y}}\ket{y}$ where $y,b\in \{0,1\}^q$ and $\tilde{y}$ is the bit string corresponding to the integer $\abs{2^{q-1} - y_{int}}$. Even though the operator {\tt HD} requires two registers, the second register will always be in the state $\ket{0^q}$ and shall be reused by uncomputing (using $HD^\dagger$) after the {\tt CMP} gate. Hence, we have not explicitly mentioned it in Algorithm~\ref{algo:qbound_fmax} and Figure~\ref{fig:qbound_fmax}. For all practical purposes, this operator can be treated as the mapping $\ket{y} \mapsto \ket{\tilde{y}}$.
    \item[{\tt CMP}:] ${\tt CMP}$ is defined as 
    ${\tt CMP} \ket{y_1}\ket{y_2}\ket{b} = \ket{y_1}\ket{y_2}\ket{b \oplus (y_1\le y_2)}$ where $y_1, y_2\in \{0,1\}^n$ and $b\in \{0,1\}$ and it simply checks if the integer corresponding to the basis state in the first register is at most that in the second register.
    \item[${\tt Cond\text{-}MAJ}_z$:] Given $k$ copies of the form $\ket{\chi^i} = \ket{x_i}\ket{c_i}$ where $x_i \in \{0,1\}^n$ and $c_i \in \{0,1\}$ and an answer register $\ket{b}$, ${\tt Cond\text{-}MAJ}_z$ flips $\ket{b}$ if $\ket{\chi^i} = \ket{z}\ket{1}$ for at least $k/2$ many indices $i$. 
\end{description}

The ${\tt EQ}$ circuit is trivial to implement, but the other three are slightly non-trivial. We have discussed their implementation details in Appendix~\ref{appendix:subroutine}.

\begin{lemma}\label{lemma:QBoundFMax}
The Algorithm {\tt QBoundFMax} makes $O\Big(\frac{1}{\epsilon\sqrt{\tau-2\epsilon}}\log(\frac{1}{\delta\tau})\log(\frac{1}{\sqrt{\delta}})\Big) = \tilde{O}(\frac{1}{\epsilon\sqrt{\tau-2\epsilon}})$ calls to the oracle and with error at most $\delta$ behaves as follows:
\begin{enumerate}
    \item if {\tt QBoundFMax} returns {\tt TRUE} then $\hat{f}^2_{max}\ge \tau - 2\epsilon$
    \item if {\tt QBoundFMax} returns {\tt FALSE} then $\hat{f}^2_{max} < \tau$.
\end{enumerate}
when given an $n-$bit Boolean function $f$ as an oracle, a threshold $\tau$, accuracy parameter $\epsilon$ and error parameter $\delta$ such that $\tau > 2\epsilon$.

Alternatively, we get that (a) if $\fmax^2 \ge \tau$ then the algorithm returns {\tt TRUE} and (b) if $\fmax^2 < \tau - 2\epsilon$ then it returns {\tt FALSE}.
\end{lemma}

\begin{proof}
Before we prove the correctness of the algorithm, we introduce a few propositions that will help us in the proof.

\begin{proposition}\label{lemma:sin_inequality}
For any two angles $\theta_1, \theta_2 \in [0,\pi]$, $\sin{\theta_1}\le \sin{\theta_2} \iff \abs{\frac{\pi}{2}-\theta_1} \ge \abs{\frac{\pi}{2}-\theta_2}$.
\end{proposition}
\begin{proof}
    The proof uses trigonometric identities and transformations.
\[
    \sin{\theta_1} \le \sin{\theta_2} \iff \cos{\Big(\frac{\pi}{2}-\theta_1\Big)} \le \cos{\Big(\frac{\pi}{2}-\theta_2\Big)} \equiv \cos\abs{\frac{\pi}{2}-\theta_1} \le \cos{\abs{\frac{\pi}{2}-\theta_2}}.
\]
Now, since $\theta_1 \in [0,\pi]$, we have $\Big(\frac{\pi}{2}-\theta_1\Big) \in [-\frac{\pi}{2},\frac{\pi}{2}]$. This in turn implies $\abs{\frac{\pi}{2}-\theta_1} \in [0,\frac{\pi}{2}]$.
    The proposition now follows from the fact that $\cos \theta$ is decreasing in the range $\theta \in [0,\tfrac{\pi}{2}]$.
\end{proof}

\begin{restatable}{proposition}{hdqprop}\label{prop:hdq}
For any two $q$-bit integers $u,v$, $$\abs{2^{q-1} - v} \le \abs{2^{q-1} - u} \iff \sin^2(\pi\frac{v}{2^q}) \ge \sin^2(\pi\frac{u}{2^q}).$$
\end{restatable}

\begin{proof}
If $z$ is a $q$-bit integer then $\pi \frac{z}{2^q} \in [0,\pi]$. Therefore, $\sin^2(\pi \frac{v}{2^q}) \le \sin^2(\pi \frac{u}{2^q})$ is equivalent to $\sin(\pi \frac{v}{2^q}) \le \sin(\pi \frac{u}{2^q})$. Using Proposition~\ref{lemma:sin_inequality}, this is equivalent to $\abs{\frac{\pi}{2}-\pi\frac{v}{2^q}} \ge \abs{\frac{\pi}{2}-\pi\frac{u}{2^q}}$ which is same as $\abs{2^{q-1} - v} \ge \abs{2^{q-1} - u}$.
\end{proof}

\begin{restatable}{proposition}{taurelprop}\label{prop:tau-rel} $\tau$ and $\tau_1$ satisfy  
$\displaystyle 0\le \tau - \epsilon - \sin^2\left(\pi\frac{\tau_1}{2^q}\right) \le \frac{2\pi}{2^q}$.
\end{restatable}
The proof of Proposition~\ref{prop:tau-rel} is given in Appendix~\ref{appendix:prop-proof}.

We analyze the algorithm in stages. Consider one of the $21\ln(\frac{1}{\delta^2\tau^2})$ independent copies. Then the state of that copy after stage-1 can be given as
$$\ket{\psi^i_1} = \sum_{x\in \{0,1\}}\hat{f}(x)\ket{x}\ket{0^n}\ket{0^l}\ket{\tau_1}\ket{0}.$$
In stage-2, on applying amplitude estimation with $R_2^i$ as the input register, $R_3^i$ as the precision register and $R_1^i$ as the register indicating the ``good state" whose amplitude should be amplified, we obtain the state of the system as,
$$\ket{\psi_2^i} = \sum_{x\in \{0,1\}^n}\hatf{x}\ket{x}\ket{\phi}\Big(\beta_{x,s}\ket{a_x}+\beta_{x,\overline{s}}\ket{E_x}\Big)\ket{\tau_1}\ket{0}.$$ 
Here $\ket{\phi}$ itself is $DJ\ket{0^n}$ [the state on which amplitude estimation happens], and
$\ket{a_x}$ is a normalized state of the form $\ket{a_x} = \gamma_{+}\ket{a_{x+}} + \gamma_{-}\ket{a_{x-}}$ that on measurement outputs $a \in \{a_{x,+}, a_{x,+}\}$ which is
an $l$-bit string that behaves as $\displaystyle\left| \sin^2\left(\frac{a\pi}{2^l}\right) - \hat{f}(x) \right| \le \tfrac{1}{2^q}$. We denote the normalized amplitude of $\ket{a_x}$ $\beta_{x,s}$ ($s$ indicates ``success'').
We denote the set $\{a_{x+}, a_{x-}\}$ by $\pmset_{a_x}$. The state $\ket{E_x}$ is the normalized error state defined as $\ket{E_x} = \sum_{a\notin \pmset_{a_x}}\gamma_{x,a}\ket{a}$. 
Notice that since the the amplitude amplification routine is called with error at most $1-\frac{8}{\pi^2}$, we have
\begin{equation}\label{eq:1}
    |\beta_{x,\overline{s}}|^2 \le 1-\tfrac{8}{\pi^2}; \mbox{ furthermore, }|\beta_{x,s}|^2 + |\beta_{x,\bar{s}}|^2=1.
\end{equation}

\indent In stage-3, the action of $\hdq{l}$ on any $l$-bit computational basis state $\ket{y}$ can be given as $$\hdq{l}\ket{y} = \ket{2^{(l-1)}-y}.$$
Meanwhile, the action of $\cmp$ on a basis state of the form $\ket{y}\ket{z}\ket{0}$ can be given as,
$$\cmp\ket{y}\ket{z}\ket{0} \xrightarrow{} \ket{y}\ket{z}\ket{y\le z}.$$
Combining these with Proposition~\ref{prop:hdq}, the action of stage-3 on a system with the basis state $\ket{y}\ket{z}\ket{0}$ can be given as
$$({\tt HD_l^{\dagger}}\otimes{\tt HD_l^{\dagger}}\otimes I)~\cmp~(\hdq{l}\otimes\hdq{l}\otimes I) \ket{y}\ket{z}\ket{0} \xrightarrow{} \ket{y}\ket{z}\ket{\mathbbm{I}\{y,z\}}$$
where $\mathbbm{1}\{y,  z\}$ is the indicator function that takes on value $1$ if $\sin^2\big(\frac{y\pi}{2^l}\big)\ge \sin^2\big(\frac{z\pi}{2^l}\big)$ and $0$ else. 
Let $\Breve{a} = \sin^2\big(\frac{a\pi}{2^l}\big)$.
We use the notation $\bre{y}$ for the expression $\sin^2(\frac{y\pi}{2^l})$. Thus, we can define $\mathbbm{1}\{y,z\}$ as $1$ if $\bre{y}\ge \bre{z}$ and $0$ otherwise.
\\

Using the above notations, we can give the transformation of $\ket{\psi_2^i}$ through stage-3 as
\begin{align*}
    \ket{\psi_2^i} &= \sum_{x\in \{0,1\}^n}\hatf{x}\ket{x}\ket{\phi}\Big(\beta_{x,s}\gamma_+\ket{a_{x,+}}+\beta_{x,s}\gamma_-\ket{a_{x,-}}+\beta_{x,\overline{s}}\ket{E_x}\Big)\ket{\tau_1}\ket{0}\\
    &\xrightarrow{\text{Stage-3}} \sum_{x\in \{0,1\}^n}\hatf{x}\ket{x}\ket{\phi}\Bigg[\beta_{x,s}\Big(\gamma_+\ket{a_{x,+}}\ket{\tau_1}\ket{\mathbbm{1}\{a_{x,+}, \tau_1\}}+\gamma_-\ket{a_{x,-}}\ket{\tau_1}\ket{\mathbbm{1}\{a_{x,-}, \tau_1\}}\Big)\\
    &\cuquad +\beta_{x,\overline{s}}\sum_{a\notin \pmset_{a_x}}\gamma_{x,a}\ket{a}\ket{\tau_1}\ket{\mathbbm{1}\{a, \tau_1\}}\Bigg]\\
    &= \sum_{x\in \{0,1\}^n}\hatf{x}\ket{x}\ket{\phi}\Bigg[\beta_{x,s}\Big(\gamma_+\ket{a_{x,+}}\ket{\tau_1}\ket{\mathbbm{1}\{a_{x,+}, \tau_1\}}+\gamma_-\ket{a_{x,-}}\ket{\tau_1}\ket{\mathbbm{1}\{a_{x,-}, \tau_1\}}\Big)\\
    &\cuquad +\beta_{x,\overline{s}}\Big(\sum_{\substack{a\notin \pmset_{a_x}\\ \bre{a}< \bre{\tau_1}}}\gamma_{x,a}\ket{a}\ket{\tau_1}\ket{0}+\sum_{\substack{a\notin \pmset_{a_x}\\ \bre{a} \ge \bre{\tau_1}}}\gamma_{x,a}\ket{a}\ket{\tau_1}\ket{1}\Big)\Bigg] = \ket{\psi^i_3}\\
\end{align*}

Now, we analyse the value of the indicator function $\mathbbm{1}\{a_x,\tau_1\}$ for $a\in \pmset_{a_x}$ under different scenarios. Recall that $\mathbbm{1}\{a,\tau_1\} = 1$ iff $\bre{a}\ge \bre{\tau}_1$.

\paragraph*{Scenario (i):}
Consider the scenario where $x$ is such that $\hat{f}(x)< \tau-2\epsilon$. Then, for any $a\in \pmset_{a_x}$, we have $\bre{a} = \sin^2\big(\frac{a\pi}{2^l}\big) \in \big[\hatf{x}-\frac{1}{2^q} , \hatf{x}+\frac{1}{2^q}\big]$. 
This gives $\bre{a} \le \tau - 2\epsilon +\frac{1}{2^q}$. Since, $q\ge \log(\frac{1}{\epsilon})+4$, we get $\frac{1}{2^q} \le \frac{\epsilon}{16}$ and hence $\bre{a} \le \tau-\epsilon-\frac{15}{16}\epsilon$. 
Using the inequality of Proposition~\ref{prop:tau-rel}, we have $\bre{a} \le \sin^2\big(\frac{\pi\tau_1}{2^l}\big)+ \frac{2\pi}{2^q} - \frac{15\epsilon}{16} \le \sin^2\big(\frac{\pi\tau_1}{2^l}\big)+ \frac{2\pi\epsilon}{16} - \frac{15\epsilon}{16} < \sin^2\big(\frac{\pi\tau_1}{2^l}\big) = \tau_1$ where the second inequality comes from the fact that $\frac{1}{2^q}\le \frac{\epsilon}{16}$.
Hence, for any $a\in \pmset_{a_x}$, we have $\mathbbm{1}\{a,\tau_1\} = 0$.

\paragraph*{Scenario (ii):}
Next, consider the scenario where $x$ is such that $\hatf{x} \ge \tau$. Again for any $a\in \pmset_{a_x}$, we have $\bre{a} \in \big[\hatf{x}-\frac{1}{2^q} , \hatf{x}+\frac{1}{2^q}\big]$.
Accordingly, $\bre{a} \ge \hatf{x}-\frac{1}{2^q}$. Since, $\hatf{x}\ge \tau$ and $\frac{1}{2^q}\le \frac{\epsilon}{16}$, we have $\bre{a} \ge \tau-\frac{\epsilon}{16} \ge \tau-\epsilon = \tau'$.
Now, we have set $\tau_1 = \left\lfloor{\frac{2^l}{\pi}\sin^{-1}(\sqrt{\tau'})}\right\rfloor \le \frac{2^l}{\pi}\sin^{-1}(\sqrt{\tau'})$, so we get $\sin^2(\frac{\pi \tau_1}{2^l}) \le \tau'$.
Hence, we have $\sin^2(\frac{\pi \tau_1}{2^l}) \le \tau' \le \bre{a} = \sin^2\big(\frac{a\pi}{2^l}\big)$, i.e, $\bre{\tau}_1 \le \bre{a}$ . This implies that for any $a\in \pmset_{a_x}$, we get $\mathbbm{1}\{a,\tau_1\} = 1$.
\\

For an analysis of stages 4 and 5, consider the following two cases.
\paragraph*{Case (i):} For all $x\in \{0,1\}^n$, $\hatf{x}< \tau-2\epsilon$.

Then, the state after stage-3 can be written as
\begin{align*}
    \ket{\psi_3^i} &= \sum_{x\in \{0,1\}^n}\hatf{x}\ket{x}\ket{\phi}\Bigg[\beta_{x,s}\ket{a_x}\ket{\tau_1}\ket{0}\\
    &\cuquad + \beta_{x,\overline{s}} \Bigg\{ \sum_{\substack{a\notin \pmset_{a_x} \\ \bre{a}<\bre{\tau_1}}}\gamma_{x,a}\ket{a}\ket{\tau_1}\ket{0} + \sum_{\substack{a\notin \pmset_{a_x} \\ \bre{a}\ge\bre{\tau_1}}}\gamma_{x,a}\ket{a}\ket{\tau_1}\ket{1} \Bigg\}\Bigg]\\
\end{align*}
The gates in stage-4 operate only on the registers $\{R^i_1, R^i_5\}_i$ (along with $R_{fo}$ and $R_{fi}$). Hence, we can ignore the $R^i_2, R^i_3$ and $R^i_4$ registers, and rewrite the residual state of $\ket{\psi_3^i}$ as
$$\ket{\chi^i} = \sum_{x\in\{0,1\}^n}\hatf{x}\ket{x}\big(\eta_{x,0}\ket{0} + \eta_{x,1}\ket{1}\big)$$
where $|\eta_{x,0}|^2 \ge |\beta_{x,s}|^2$. 
Since $|\beta_{x,s}|^2 + |\beta_{x,\bar{s}}|^2=1$, $|\eta_{x,1}|^2 \le |\beta_{x,\bar{s}}|^2$.

Now, in stage-4, for every basis state $\{ \ket{x} ~:~ x\in\{0,1\}^n\}$ of $R_{fi}$, we perform a conditional majority of the $21\ln(\frac{1}{\delta^2\tau^2})$ copies of $R_5^i$ conditioned on $R_1^i$ being $\ket{x}$, and we store the result in $R_{fo}$. Suppose $k$ denotes the number of independent copies.
To compute the probability that $R_{fo}$ is $\ket{1}$ when $R_{fi}$ is in $\ket{x}$, observe that this event happens when at least $k/2$ of the $\{R^i_1, R^i_5\}_i$ pairs of registers are in the state $\ket{x}\ket{1}$. For any $i$, the probability that $R^i_1 R^i_5$ of $\ket{\chi^i}$ is in the state $\ket{x}\ket{1}$ is $|\hat{f}(x) \eta_{x,1}|^2$ which is less than $|\hat{f}(x)|^2 |\beta_{x,\bar{s}}|^2 \le |\beta_{x,\bar{s}}|^2$ that can be upper bounded by $1 - \tfrac{8}{\pi^2}$ (using Equation~\ref{eq:1}).

Then, using Chernoff bound\footnote{
Given a coin whose probability of head is $p > \tfrac{1}{2}$, Chernoff's bound says that the probability that tail is observed in at least $\tfrac{n}{2}$ trials out of $n$ Bernoulli trials is upper bounded by $\exp{-\frac{1}{2p}n(p-\frac{1}{2})^2}$. In our case, $1 \ge p > \tfrac{8}{\pi^2}$, so $n \ge \tfrac{2}{(\frac{8}{\pi^2} - \frac{1}{2})^2} \ln \tfrac{1}{b}$ is sufficient for error at most $b$.}
it is straight forward to see the following relation for each $x\in \{0,1\}^n$:
$$Pr\Big[R_{fo}=\ket{1}\Big|R_{fi}=\ket{x}\Big] \le \delta^2\tau^2.$$ Since, in this case, the above relation holds true for all $x\in \{0,1\}^n$, we have that $Pr\big[R_{fo}=\ket{1}\big] \le \delta^2\tau^2$.

\indent In stage-5, we perform amplitude amplification on $R_{fo}$ with $\ket{1}$ being the ``good" state using the fixed-point amplitude amplification algorithm (FPAA)~\cite{fixedpointaa}. $\frac{1}{\sqrt{\lambda}}\log(\frac{2}{\sqrt{\delta}})$ calls to the oracle are necessary and sufficient for FPAA to output a good state with probability at least $1-\delta$ where $\lambda$ is the probability of the good state prior to amplification.\footnote{We can replace FPAA by vanilla amplitude amplification where success probability is unknown but a lower-bound is known. A similar analysis will follow.} 
Thus, the number of iterations required to amplify the probability of a state from some probability that is at most $\delta^2\tau^2$ to $\delta$ is $\Omega(\frac{1}{\tau\delta} \log \tfrac{2}{\sqrt{\delta}}) \gg \frac{1}{\sqrt{\tau-2\epsilon}}\log(\frac{2}{\sqrt{\delta}})$ which is the number of amplifications that the algorithm performs. Hence, the error will be (much) larger than $\delta$, or in other words, the probability of obtaining $R_{fo}$ as $\ket{0}$ on measurement after stage-5 is at least $1-\delta$.

\paragraph*{Case (ii):} Now, let there exist some $z\in \{0,1\}^n$ such that $\hatf{z}\ge \tau$.

Let $G$ be the set of all such $z$, i.e, $G=\{z\in \{0,1\}^n : \hatf{z}\ge \tau\}$.
Then the state after stage-3 can be written as
\begin{align*}
    \ket{\psi_3^i} &= \sum_{x\in \{0,1\}^n}\hatf{x}\ket{x}\ket{\phi}\Bigg[\beta_{x,s}\ket{a_x}\ket{\tau_1}\ket{\mathbbm{1}\{a_x,\tau_1\}}\\
    &\cuquad + \beta_{x,\overline{s}} \Bigg\{ \sum_{\substack{a\notin \pmset_{a_x} \\ \bre{a}<\bre{\tau_1}}}\gamma_{x,a}\ket{a}\ket{\tau_1}\ket{0} + \sum_{\substack{a\notin \pmset_{a_x} \\ \bre{a}\ge\bre{\tau_1}}}\gamma_{x,a}\ket{a}\ket{\tau_1}\ket{1} \Bigg\}\Bigg]\\
    & = \sum_{x\in G}\hatf{x}\ket{x}\ket{\phi}\Bigg[\beta_{x,s}\ket{a_x}\ket{\tau_1}\ket{1}\\
    &\cuquad +\beta_{x,\overline{s}} \Bigg\{ \sum_{\substack{a\notin \pmset_{a_x} \\ \bre{a}<\bre{\tau_1}}}\gamma_{x,a}\ket{a}\ket{\tau_1}\ket{0} + \sum_{\substack{a\notin \pmset_{a_x} \\ \bre{a}\ge\bre{\tau_1}}}\gamma_{x,a}\ket{a}\ket{\tau_1}\ket{1} \Bigg\}\Bigg]\\
    & + \sum_{x\notin G}\hatf{x}\ket{x}\ket{\phi}\Bigg[\beta_{x,s}\ket{a_x}\ket{\tau_1}\ket{0}\\
    &\cuquad +\beta_{x,\overline{s}} \Bigg\{ \sum_{\substack{a\notin \pmset_{a_x} \\ \bre{a}<\bre{\tau_1}}}\gamma_{x,a}\ket{a}\ket{\tau_1}\ket{0} + \sum_{\substack{a\notin \pmset_{a_x} \\ \bre{a}\ge\bre{\tau_1}}}\gamma_{x,a}\ket{a}\ket{\tau_1}\ket{1} \Bigg\}\Bigg]
\end{align*}

As before, if we trace out the registers $R^i_2, R^i_3$ and $R^i_4$ in $\ket{\psi^i_3}$, the residual state can be expressed as
$$\ket{\chi^i} = \sum_{x\in G}\hatf{x}\ket{x}\big(\eta_{x,1}\ket{1} + \eta_{x,0}\ket{0}\big) + \sum_{x\notin G}\hatf{x}\ket{x}\big(\eta_{x,0}\ket{0}+\eta_{x,1}\ket{1}\big)$$ where $|\eta_{x,1}|^2 \ge |\beta_{x,s}|^2$ and $|\eta_{x,0}|^2 \le |\beta_{x,\overline{s}}|^2$ for $x\in G$, and $|\eta_{x,0}|^2 \ge |\beta_{x,s}|^2$ and $|\eta_{x,1}|^2 \le |\beta_{x,\overline{s}}|^2$ for $x\notin G$,

Next, for every $x\in\{0,1\}^n$, conditioned on all the $R^i_1$ registers and $R_{fi}$ being in state $\ket{x}$, we perform a conditional majority over all the $R^i_5$ registers and store the output in $R_{fo}$.
Then, using Chernoff bounds as in case(i), we get that for any $x\in G$, $$Pr\Big[R_{fo}=\ket{1}\Big|R_{fi}=\ket{x}\Big] \ge 1-\delta^2\tau^2 \ge 1-\delta,$$ and for any $x\notin G$ we have, $$Pr\Big[R_{fo}=\ket{1}\Big|R_{fi}=\ket{x}\Big] \le \delta^2 \tau^2 < \delta.$$

Therefore, the overall probability of obtaining $\ket{1}$ in $R_{fo}$ after stage-4 can be expressed as
$$Pr\Big[R_{fo}=\ket{1}\Big] \ge \sum_{x\in G}|\hatf{x}|^2\cdot (1-\delta) \ge \tau(1-\delta) \ge \tau/2$$ under the reasonable assumption that the target error probability $\delta < \tfrac{1}{2}$.

Thus on amplifying the amplitude of the state $\ket{1}$ in $R_{fo}$ for $\frac{1}{\sqrt{\tau-2\epsilon}}\log(\frac{2}{\sqrt{\delta}})$ iterations using fixed point amplitude amplification, we observe $\ket{1}$ in $R_{fo}$ with probability at least $1-\delta$ as required.
\\

Now, we evaluate the query complexity of the algorithm. It is straightforward to observe that the number of calls made by amplitude estimation in {\tt QBoundFMax} is $O\big(2^l\big)$ = $O\big(\frac{1}{\epsilon}\big)$.
The amplitude estimation subroutine is implemented on $21\ln(\frac{1}{\delta^2\tau^2}) = O\big(\log(\frac{1}{\delta\tau})\big)$ many independent copies.
Hence the query complexity at the end of stage-2 is $O\big(\frac{1}{\epsilon}\log(\frac{1}{\delta\tau})\big)$.
In the last stage, amplitude amplification is performed $O\Big(\frac{1}{\sqrt{\tau - 2\epsilon}}\log(\frac{1}{\sqrt{\delta}})\Big)$ iterations. Hence, the algorithm {\tt QBoundFMax} makes $O\big(\frac{1}{\epsilon}\log(\frac{1}{\delta\tau})\big)\cdot O\Big(\frac{1}{\sqrt{\tau - 2\epsilon}}\log(\frac{1}{\sqrt{\delta}})\Big) = \tilde{O}\big(\frac{1}{\epsilon\sqrt{\tau - 2\epsilon}}\big)$ queries to the oracle in total.

\end{proof}

\subsection{Fine tuning of interval search}\label{sec:high}

Earlier we saw how to obtain a small interval $J$ that contains
$\fmax^2$ with
high probability. However, there may be a requirement to
fine-tune this estimation.

Suppose $f$ is linear, i.e., $\nlf=0$. For such a function $\fmax^2=1$; 
however, due to the nature of Algorithm~\ref{algo:intervalsearch} we will get the
interval $[1-\epsilon,1]$for $\fmax^2$, and hence an interval for $\nlf$. We feel that a
non-linearity estimation algorithm should be able to clearly identify a linear
function instead of presenting approximate values close to 0.

Consider the other extreme of Bent functions with the largest
non-linearity; these would have $\fmax^2 = \frac{1}{2^n}$. However, our
Algorithm~\ref{algo:intervalsearch} will, most-likely, return the interval
$\approx [\tfrac{1}{2^n},\epsilon)$. We wonder if it is possible to obtain an even tighter interval.

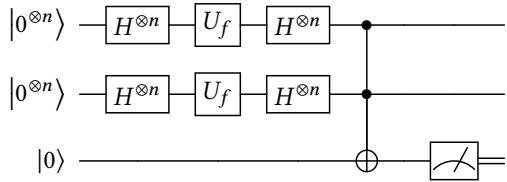
\begin{figure}[!h]
	    $$
            \Qcircuit @C=1.0em @R=1.0em {
            \lstick{\ket{0^{\otimes n}}} & \gate{H^{\otimes n}} & \gate{U_f} & \gate{H^{\otimes n}} & \ctrl{2} &\qw &\qw &\qw\\
            \lstick{\ket{0^{\otimes n}}} & \gate{H^{\otimes n}} & \gate{U_f} & \gate{H^{\otimes n}} & \ctrl{1} &\qw &\qw &\qw\\
            \lstick{\ket{0}} & \qw & \qw & \qw & \targ & \qw &\meter &\cw\\
            }
        $$
    \caption{Circuit for estimating $\sum_x \hat{f}^4(x)$\label{fig:est_qalgo}}
\end{figure}

For handling these extreme values of $\fmax^2$ consider the quantum
circuit illustrated in Figure~\ref{fig:est_qalgo}.
Three registers $R_1$, $R_2$ and $R_3$ are initialized to $\ket{0^{\otimes n}}$, $\ket{0^{\otimes
n}}$ and $\ket{0}$, respectively.
Then it applies the Deutsch-Jozsa circuit independently on $R_1$ and
$R_2$, to obtain the state $\sum_{x}\sum_{y}\hat{f}(x)\hat{f}(y)\ket{x}\ket{y}\ket{0}$.
Finally, the circuit flips the state of $R_3$ if $x = y$ to obtain the
state $\sum_{x,y:x=y}\hat{f}^2(x)\ket{x}\ket{x}\ket{1} +
\sum_{x,y:x\neq y}\hat{f}(x)\hat{f}(y)\ket{x}\ket{y}\ket{0}$.
Now $R_3$ is measured in the standard basis.
The probability of observing $\ket{1}$ in $R_3$ is
$$p = \sum_x{\hat{f}^4}\le\hat{f}^2_{max}\sum_x{\hat{f}^2}=\hat{f}_{max}^2
\quad\mbox{(using Parseval's equality)}.$$

At this point amplitude estimation can be used to estimate this probability
with { any desired} additive error $\epsilon'$, and with low error probability
denoted by $\delta'$.
Denoting this estimate by $p^*$, with probability at least $1-\delta'$ it satisfies $p^* -
\epsilon' \le p \le p^* + \epsilon'$, which implies that $\fmax^2 \ge p^* -
\epsilon'$. This is used to get a tighter lower bound for $\fmax^2$.
The value of $\delta'$ \st{will be decided in \mbox{Section~\ref{sec:nonlinalgo}}} { is set to the same $\delta$ that is used in IntervalSearch}.
A tight estimate for $\fmax^2$ is computed based on the interval $J$ obtained
from IntervalSearch.

If $J$ is the rightmost interval, then we want to primarily detect if
$\fmax^2=1$. We will use $\epsilon'=\epsilon$. Note that amplitude estimation will return 1 in this case without
fail. Thus, if $p^*=1$, then the interval $[1,1]$ is returned. Otherwise, the
left boundary of $J$ can be tightened to $p^* - \epsilon$, provided it is larger
than the original left boundary of $J$. Clearly, we get an interval of length at
most $\epsilon$. The number of calls to the circuit, and hence to $U_f$ will be
$O(\frac{1}{\epsilon}\log \frac{1}{\delta})$.

If $J$ is the leftmost interval, then we want to primarily get a lower bound on
$\fmax^2$ larger than $\tfrac{1}{2^n}$. For this reason, we will use
$\epsilon'=\epsilon^{3/2}$ and employ the improved amplitude estimation proposed
by Montanaro~\cite{montanaro2015quantum}. Observe that the output probability of the
circuit is $p=\sum_x \hat{f}^4(x)$; so, its variance can be computed as
$$p(1-p) \le \fmax^2(1-\sum_x \hat{f}^4(x)) \le \fmax^2 \le \epsilon,$$
where the last inequality is implied by the fact that IntervalSearch returned
$J$ as the last interval whose right boundary ($\tfrac{1}{2^k}$) is less than
$\epsilon$. Montanaro showed that if variance of an output is bounded, then a
better estimation method exists for additive errors; in our case, the number of calls to the
circuit, and so also to $U_f$ will be
$O(\tfrac{\sqrt{\epsilon}}{\epsilon^{3/2}} \log\frac{1}{\delta}) =
O(\tfrac{1}{\epsilon} \log \frac{1}{\delta})$. The estimation procedure will
return an estimate $p^*$. If $p^* \le \epsilon^{3/2}$ then we will can safely
use $\tfrac{1}{2^n}$ as the left boundary of $J$. Otherwise, we can increase the
left boundary of $J$ to $p^*-\epsilon^{3/2}$. In both the cases, the length of
$J$ remains at most $\epsilon$.

To summarise, we saw above how to make $O(\frac{1}{\epsilon} \log \frac{1}{\delta})$
queries to $U_f$ and return an estimate for $\fmax^2$ that is at most $\epsilon$
away from the actual value. Moreover, if $f$ is linear, it correctly identifies
that $\fmax^2=1$.

\section{Non-linearity estimation}\label{sec:nonlinalgo}

Now we combine the earlier results to present our algorithm for
estimating $\nlf$.
	
\begin{theorem}\label{thm:nonlinalgo}
Given an oracle $U_f$ to an $n$-bit  function $f$, an additive accuracy
$\lambda$ and an error $\delta$, there is a quantum algorithm that outputs an
interval $I$ of length at most $\lambda$ such that the normalised
nonlinearity of the function $f$, denoted by $\nlf$, belongs to $I$ with
probability at least $1-\delta$. The algorithm makes $\Ot \left(
\frac{1}{\lambda^3})\right)$ queries to $U_f$. If $\nlf=0$ then the
algorithm always outputs the correct non-linearity.
\end{theorem}

The algorithm simply calls {\tt IntervalSearch} (along with fine-tuning
described in Section~\ref{sec:high}) 
using accuracy $\epsilon = 2d$ (for some
$d$ to be decided below) and using the error probability \st{$\delta$} { $\delta/2$. The combined probability of error, from IntervalSearch and fine-tuning, remains bounded by $\delta$.} Let $J=[c - d, c+d)$ be the interval
returned by {\tt IntervalSearch}. If $J=[1,1]$ then the algorithm shall output $J$, else $I=\Big(
\frac{1}{2}(1-\sqrt{c})-\frac{\sqrt{d}}{2},\frac{1}{2}(1-\sqrt{c})+\frac{\sqrt{d}}{2}
\Big]$. We prove the correctness of this algorithm below.

\begin{proof}
$\nlf=0$ is equivalent to $\fmax^2=1$ and in that case, {\tt
IntervalSearch} after fine-tuning returns $[1,1]$ without fail.
Thus, this case is correctly handled.

For non-linear functions, suppose {\tt IntervalSearch} returns $J = [c-d, c+d)$.
{\tt IntervalSearch} guarantees that $\fmax^2 \in J$. Then, $\fmax \in J' = \Big[\sqrt{c-d}, \sqrt{c+d}\Big)$ and
so, $\nlf \in I' =
\Big(\frac{1}{2}(1-\sqrt{c+d}),\frac{1}{2}(1-\sqrt{c-d})\Big]$.
From the fact that $\sqrt{a} - \sqrt{b} \le \sqrt{a-b}$ and $\sqrt{a} +
\sqrt{b} \ge \sqrt{a+b}$, we can say that $I'$ is contained in $I = \Big(
\frac{1}{2}(1-\sqrt{c}-\sqrt{d}),\frac{1}{2}(1-\sqrt{c}+\sqrt{d}) \Big] =
\Big(
\frac{1}{2}(1-\sqrt{c})-\frac{\sqrt{d}}{2},\frac{1}{2}(1-\sqrt{c})+\frac{\sqrt{d}}{2}
\Big]$.

The length of $I$ is $\sqrt{d}$ that we want to be $\lambda$. So, we set
$d=\lambda^2$.
From Theorem~\ref{thm:int_search}, we know that the complexity of {\tt IntervalSearch} when using {\tt QBoundFMax} is $\tilde{O}\left(\frac{1}{\epsilon^{3/2}} \right)$.  
Here, we have $\epsilon = 2d = 2\lambda^2$.
Hence, the number of $U_f$ queries in {\tt
IntervalSearch} while using {\tt QBoundFMax} is upper bounded by $\Ot\left( \frac{1}{\lambda^3} \right)$.
\end{proof}

\begin{theorem}\label{thm:cnonlinalgo}
There is a classical algorithm that makes $\Ot(\frac{n}{\lambda^6})$ queries to $U_f$ and outputs an
interval $I$ of length at most $\lambda$ such that $\nlf$ belongs to $I$ with constant probability.
\end{theorem}

The algorithm and the proof are exactly as in Theorem~\ref{thm:nonlinalgo} but using {\tt CBoundFMax} and without the fine-tuning subroutine.

\section{Lower Bounds on Non-linearity}\label{sec:lower_bound}
Now we present our query complexity lower bounds for a non-linearity separation problem. We employ the well-known quantum adversary method proposed by Ambainis~\cite{qam} which is stated in the form of Theorem~\ref{thm:qa_method}.
\begin{theorem}
\label{thm:qa_method}
Let $F$ be a $p$-bit Boolean function and $X$ and $Y$ be two sets of inputs such that $F(x)\neq F(y)$ for any $x\in X$ and $y\in Y$. Let $R\subseteq X\times Y$ be a relation such that 
\begin{enumerate}
    \item for every $x \in X$, $\exists$ at least $m$ different $y \in Y$ such that $(x,y)\in R$.
    \item for every $y \in Y$, $\exists$ at least $ m'$ different $x \in X$ such that $(x,y)\in R$.
    \item for every $x \in X$ and $i \in \{1,...,p\}$, $\exists$ at most $l$ different $y \in Y$ such that $x_i \neq y_i$ and $(x,y)\in R$.
    \item for every $y \in Y$ and $i \in \{1,...,p\}$, $\exists$ at most $l'$ different $x \in X$ such that $x_i \neq y_i$ and $(x,y)\in R$.
\end{enumerate}
Then any quantum algorithm uses $\Omega\Big(\sqrt{\frac{m\cdot m'}{l\cdot l'}}\Big)$ queries to compute $F$ { on $X \bigcup Y$}.
\end{theorem}


Consider the following $\hat{f}_{max}$ decision problem: Given a Boolean function $f$ with a promise that $\hat{f}_{max} = 1$ or $\hat{f}_{max} = 1-4\lambda$, decide the case of $f$.
\st{Mathematically,} { Using the notations of Theorem~\ref{thm:qa_method}, consider a Boolean function $F:\{0,1\}^{2^n}\rightarrow \{0,1\}$ that takes as input a truth-table of an $n$-bit Boolean function encoded as a $2^n$-bit binary string and outputs 1 iff the encoded function, say $f()$, satisfies $\hat{f}_{max}=1$. The $\fmax$ decision problem is to compute $F()$ on $X \bigcup Y$ where $X$ contains some $n$-bit functions with $\fmax=1$ and $Y$ contains some $n$-bit functions with $\fmax=1-4\lambda$.
}
\st{$F:\{0,1\}^{2^n}\rightarrow \{0,1\}$ be the function such that $F(x) = 0$ if $\hat{f_x}_{max} = 1$ and $F(x) = 1$ if $\hat{f_x}_{max} = 1 - 4\lambda$ where $f_x$ is the $n$-bit Boolean function that corresponds to the truth table $x$. Then, given a Boolean function $f_y$ with the promise, the $\hat{f}_{max}$ decision problem is to decide if $F(y)=0$ or $F(y)=1$.}
Note that any algorithm that solves non-linearity estimation with $\lambda$ accuracy also computes $\fmax$ with $2\lambda$ accuracy and, thus, can solve the above decision problem.
So, a lower bound on the $\hat{f}_{max}$ decision problem naturally is a lower bound on the problem of non-linearity estimation. 

For the case of $\lambda > \frac{1}{8}$, we have a constant time quantum algorithm for non-linearity estimation and so the lower bound is $\Omega(1)$.
So suppose that $\lambda \le \frac{1}{8}$. Consider the following sets $X = \{0^{2^n}\}$ and $Y = \{y \in \{0,1\}^n : wt(y) = 2\lambda \cdot 2^n\}$ where $wt()$ denotes the Hamming weight.
Notice that deciding if an $n$-bit Boolean function $g$ belongs to $X$ or $Y$ given that $g$ belongs to one of them conforms to the generalized Grover search problem. We present the lower bound of this problem below.

\begin{lemma}\label{lemma:dist_x_y}
Any quantum algorithm uses $\Omega(\frac{1}{\sqrt{\lambda}})$ queries to the oracle of the given function $g$ to decide if $g\in X$ or $g\in Y$ given the promise that $g$ belongs to one of them.
\end{lemma}
\begin{proof}
Consider the sets $X$ and $Y$. Let the relation $R$ be $R = X \times Y$. 
Also let $\tilde{\lambda} = 2\lambda\cdot 2^n$. Then for every $x\in X$, $\exists$ exactly $\binom{2^n}{\tilde{\lambda}}$ different $y\in Y$ such that $(x,y)\in R$ and for every $y\in Y$, $\exists$ exactly $1$ $x\in X$ such that $(x,y)\in R$.
So we have $m = \binom{2^n}{\tilde{\lambda}}$ and $m' = 1$.
Now, for any $x\in X$ and $i \in \{1,2,\cdots,2^n\}$, fix $y_i = 1$. Naturally, $x_i \neq y_i$. There are exactly $\binom{2^n-1}{\tilde{\lambda}-1}$ different $y\in Y$ such that $y_i = 1$ for any $i$. Hence, we have $l = \binom{2^n-1}{\tilde{\lambda}-1}$.
On the other side it is trivial that $l' = 1$ since there is only a single element in $X$.
Hence, from Theorem~\ref{thm:qa_method}, we obtain our lower bound as $\Omega(\frac{1}{\sqrt{\lambda}})$ queries.
\end{proof}

\emph{Note:} Since the problem in Lemma~\ref{lemma:dist_x_y} is a version of the unstructured search problem, the lower bound in Lemma~\ref{lemma:dist_x_y} is also a direct consequence of the optimality of the generalized Grover's search~\cite{general_grover_lb}.

\begin{theorem}\label{thm:lb_fhat_decide1}
Any quantum algorithm uses $\Omega(\frac{1}{\sqrt{\lambda}})$ queries to the oracle of the given function $f$ to decide if $\hat{f}_{max} = 1$ or $\hat{f}_{max} = 1-4\lambda$ given the promise that it is either of the cases.
\end{theorem}

\begin{proof}
Consider the set $X$ and $Y$ as defined earlier. We know that for any $a\in \{0,1\}^n$, $\hat{f}(a)$ can be written as $\frac{1}{2^n}(\lvert\{x: f(x)=a\cdot x\}\rvert - \lvert\{x: f(x)\neq a\cdot x\}\rvert)$. Now, since $\lambda \le \frac{1}{8}$, we have $\hat{f_y}(0^n) = 1-4\lambda \ge \frac{1}{2}$ for any $y\in Y$ where $y$ represents the truth table of $f_y$.
Next, as for any $a\neq 0^n$, $f_y(x) = a\cdot x$ for at most $(\frac{1}{2}+2\lambda)2^n$ many $x$'s. 
Hence, we have $\hat{f_y}(a) \le 4\lambda \le \frac{1}{2}$ for any $y\in Y$.
Hence for any $y\in Y$, $(\hat{f_y})_{max}$ occurs at $0^{n}$ and $(\hat{f_y})_{max} = 1-4\lambda$.
Now, for $x\in X$, we have $f_x(z) = 0$ for all $z\in \{0,1\}^n$. Hence, $\hat{f_x}(0^n)=1$ and $\hat{f_x}(a) = 0$ for any $a\neq 0$. So for $x\in X$, $(\hat{f_x})_{max}$ occurs at $0^n$ and $(\hat{f_x})_{max} = 1$.

Thus, the problem of deciding if a given function $f$ belongs to $X$ or $Y$ can be reduced to deciding if for the function $f$, $\hat{f}_{max}=1$ or $\hat{f}_{max}=1-4\lambda$. 
So, using Lemma~\ref{lemma:dist_x_y}, we obtain that any quantum algorithm uses $\Omega(\frac{1}{\sqrt{\lambda}})$ queries to the oracle of the given function $f$ to decide if $\hat{f}_{max} = 1$ or $\hat{f}_{max} = 1-4\lambda$ given the promise that it is either of the cases.
\end{proof}

Note that the lower bound in Theorem~\ref{thm:lb_fhat_decide1} indicates the difficulty in separating functions that are linear and $4\lambda$-close to linear. We also obtain the same lower bound separating functions that are bent and are $4\lambda$-close to bent. Recall that a function $f$ is called a bent function if $\abs{\hat{f}(a)} = \frac{1}{\sqrt{2^n}}$ for all $a\in \{0,1\}^n$.
Therefore, for such a function, if $\hat{f}(0^n) = \frac{1}{\sqrt{2^n}}$ then $wt(x_f) = \frac{1}{2}(2^n - \sqrt{2^n})$ where $x_f$ represents the truth table of $f$.

Characterising functions that are $4\lambda$-close to bent requires a little bit of work, and the following proposition does the heavy-lifting.

\begin{proposition}
    Let $f$ be some $n$-bit function such that $wt(x_f) \ge k$ and a Boolean function $g$ be obtained from $f$ by flipping $k$ bits in $x_f$ from 1 to 0. Then, $\hat{g}(0^n) = \hat{f}(0^n) + k\frac{2}{2^n}$, and for any $a \not= 0^n$, $\hat{g}(a) = \hat{f}(a) + \sum_{i=1}^k \pm \tfrac{2}{2^n}$.
\end{proposition}

\begin{proof}
    We will actually prove the proposition for $k=1$, i.e., $g$ is obtained from $f$ by flipping the value at some point, say $y$; so $f(y)=1$ was flipped to $g(y)=0$. The general case can be applied by repeatedly applying this case, in turn, to the $g$ that is obtained.
    
    For any $a \in \{0,1\}^n$, define $S^0_a(f)$ as the number of elements in the set $\{ x ~:~ f(x) = a\cdot x\}$ and $S^1_a(f)$ as the number of elements in the set $\{ x ~:~ f(x) \not=a \cdot x\}$. When $a=0^n$, $S^1_a(f)=wt(f)$ and $S^0_a(f)=2^n-wt(f)$. In general, $\hat{f}(a)=\tfrac{1}{2^n}(S^0_a(f) - S^1_a(f))$.

    Now, $g$ is obtained by flipping the value of $f(y)=1$ to $g(y)=0$, i.e., $wt(x_g) = wt(x_f)-1$. First, consider the case of $a=0^n$. For this $a$, $S^1_a(g)=S^1_a(f)-1$, $S^0_a(g) = S^1_a(f)+1$, and $\hat{g}(0^n) = \hat{f}(0^n) + \tfrac{2}{2^n}$. This proves the first claim of the proposition.

    Now we prove the case of general $a \not= 0^n$. There are two possibilities here. On one hand, if $a$ is such that $f(y)=a \cdot y$, then $S^0_a(g)=S^0_a(f)-1$ and $S^1_a(g) = S^1_a(f)+1$, and thus, $\hat{g}(a) = \hat{f}(a) - \frac{2}{2^n}$. On the other hand, if $a$ is such that $f(y) = 1 \xor a \cdot y$, then a similar argument shows that $\hat{g}(a) = \hat{f}(a) + \tfrac{2}{2^n}$.
    Combining both the cases we get that $\hat{g}(a) = \hat{f}(a) \pm \frac{2}{2^n}$ as required.
\end{proof}

If we choose bent functions such that $\hat{f}(0^n) = \frac{1}{\sqrt{2^n}}$, we always have $\hat{f}(0^n)\ge \hat{f}(a)$ for any $a$. Moreover, if $g$ is obtained by flipping some bits of $x_f$ from 1 to 0, then from the above proposition we get that $\hat{g}(0^n) = \hat{f}(0^n) + \tfrac{2k}{2^n} \ge \hat{f}(0^n) + \sum_{i=1}^k \pm \frac{2}{2^n} \ge \hat{f}(a) + \sum_{i=1}^k \pm \frac{2}{2^n} = \hat{g}(a)$ for any $a$. That is, $\hat{g}_{max} = \hat{g}(0^n)$. Clearly, if exactly $2\lambda\cdot 2^n$ $1s$ in $x_f$ are flipped to $0$, then $\hat{g}_{max}$ occurs at $0^n$ and $\hat{g}(0^n) = \frac{1}{\sqrt{2^n}} + 4\lambda$.

\begin{theorem}\label{thm:lb_fhat_decide2}
Any quantum algorithm uses $\Omega(\frac{1}{\sqrt{\lambda}})$ queries to the oracle of the given function $f$ to decide if $\hat{f}_{max} = \frac{1}{\sqrt{2^n}}$ or $\hat{f}_{max} = \frac{1}{\sqrt{2^n}}+4\lambda$ given the promise that it is either of the cases.
\end{theorem}

\begin{proof}
Let $x$ be the truth table string of some bent function $f_x$ such that $\hat{f_x}(0^n) = \frac{1}{\sqrt{2^n}}$. Then we know $wt(x) = \frac{1}{2}(2^n - \sqrt{2^n}) = k$.
Let $X = \{x\}$ and $\tilde{\lambda} = 2\cdot \lambda \cdot 2^n$. Now let $Y$ be the set that contains all strings $y$ which can be obtained by flipping exactly $\tilde{\lambda}$ number of $1s$ to $0$.
Then for any $y\in Y$, $wt(y) = \frac{1}{2}(2^n - \sqrt{2^n}) - \tilde{\lambda}$. Let $R = X \times Y$.
Now, for $x\in X$, we can see that there are exactly $\binom{k}{\tilde{\lambda}}$ number of $y \in Y$ such that $(x,y)\in R$ and so $m = \binom{k}{\tilde{\lambda}}$.
Similarly, we can see that for any $i \in \{1,\cdots,2^n\}$ there are at most $l = \binom{k-1}{\tilde{\lambda}-1}$ $y\in Y$ such that $x_i\neq y_i$ and $(x,y)\in R$.
Since the set $X$ contains only a single element, we have $m' = 1$ and $l' = 1$.
    So we have $\frac{m\cdot m'}{l \cdot l'} = \frac{k}{\tilde{\lambda}} = \frac{2^n - \sqrt{2^n}}{2\tilde{\lambda}} \ge \xcancel{\frac{2^n}{2\tilde{\lambda}}} \quad { \frac{2^n/2}{2\tilde{\lambda}}} $ = \xcancel{$\frac{1}{4\lambda}$} { $\frac{1}{8\lambda}$}.
Thus we have the lower bound as $\Omega(\frac{1}{\sqrt{\lambda}})$.

\end{proof}

These results lead to our required quantum lower bound.

\begin{theorem}\label{cor:nonlin_lower}
Any quantum algorithm uses $\Omega(\frac{1}{\sqrt{\lambda}})$ queries to estimate the non-linearity of a given function with $\lambda$ accuracy.
\end{theorem}

Next we show the classical randomised complexity of the non-linearity estimation. The adversary method has been extended to randomised algorithms as well which is what we use. Using a theorem given by Laplante et al.~\cite{quantum_random_lower} in which we use the values obtained in the proof of Theorem~\ref{cor:nonlin_lower}, we obtain the following result. (Complete proof is given in Appendix~\ref{appendix:rlb}.)

\begin{theorem}\label{thm:rlb}
Any classical randomised algorithm uses $\Omega(\frac{1}{\lambda})$ queries to estimate the non-linearity of a given function with $\lambda$ accuracy.
\end{theorem}

\section{Conclusion}\label{sec:conclusion}

In this paper we looked at the problem of estimating, within additive accuracy $\lambda$, the non-linearity of a Boolean function given to us as a black-box. We
devised a quantum strategy that makes $\Ot(\frac{1}{\lambda^3})$ queries to the
black-box in the worst-case and is independent of the size of the function. In contrast, our classical randomised algorithm makes linear in $n$ many queries apart from depending on $\lambda$ as $\frac{1}{\lambda^6}$.
We proved a lower bound of $\Omega(\frac{1}{\sqrt{\lambda}})$ on the quantum query complexity and a lower bound of $\Omega(\frac{1}{\lambda})$ on the classical randomised query complexity of the non-linearity estimation problem. We conclude with a conjecture that the quantum query complexity of non-linearity estimation is $\Theta(\frac{1}{\sqrt{\lambda}})$ --- that will allow us to construct a sub-exponential query quantum algorithm for exactly computing $\nlf$ using only a few additional qubits.


\newpage
\bibliographystyle{ACM-Reference-Format}
\bibliography{refs}
\newpage
\appendix

\section{Amplitude estimationn and amplification}\label{appendix:ampest}

In this section, we provide details on the quantum amplitude estimation and amplitude amplification subroutines that are used as part of our algorithms.

\subsection{Amplitude Estimation}

The amplitude estimation problem is the problem of estimating the probability $p$ of an observation upon
measuring the final state of a quantum circuit $A$. 
Let $k$ and $m$ be some parameters that we shall fix later.
A quantum amplitude estimation algorithm (say, named as $AmpEst$) was proposed by Brassard et
al.~\cite{brassard2002quantum}
that acts on two registers of $m$ and $n$ qubits, makes
$2^m$ calls to controlled-$A$ and outputs a $\tilde{p} \in [0,1]$ that is a good approximation of $p$ in the following sense.
\begin{theorem}\label{thm:amp_est}
    The $AmpEst$ algorithm returns an estimate $\tilde{p}$ that has a confidence
    interval $|p-\tilde{p}| \le 2\pi k \frac{\sqrt{p(1-p)}}{2^m} +
    \pi^2 \frac{k^2}{2^{2m}}$ with probability at least $\frac{8}{\pi^2}$ if
    $k=1$ and with probability at least $1-\frac{1}{2(k-1)}$ if $k \ge 2$. If
    $p=0$ or 1 then $\tilde{p}=p$ with certainty.
\end{theorem}


The $AmpEst$ algorithm can be used to estimate $p$ with desired accuracy and
error.
The following corollary is obtained directly from the above theorem.
\begin{corollary}\label{cor:amp_est_our_form}
    The amplitude estimation algorithm returns an estimate $\tilde{p}$ that has a confidence
    interval $|p-\tilde{p}| \le \frac{1}{2^q}$ with probability at least $\frac{8}{\pi^2}$ using $q+3$ qubits and $2^{q+3}-1$ queries. If
    $p=0$ or 1 then $\tilde{p}=p$ with certainty.
\end{corollary}
\begin{proof}
Set $k=1$ in Theorem~\ref{thm:amp_est}. Since $p\le 1$, we get $\sqrt{p(1-p)} \le \frac{1}{2}$. Then we have
$$2\pi k \frac{\sqrt{p(1-p)}}{2^m} + \pi^2 \frac{k^2}{2^{2m}} \le 2\pi\frac{1}{2\cdot 2^m} + \pi^2\frac{1}{2^{2m}} \le \frac{\pi}{2^m} + \frac{\pi^2}{2^{2m}}\le \frac{2\pi}{2^m} \le \frac{8}{2^m} = \frac{1}{2^{m-3}}.$$
The last inequality follows from the fact that $\frac{\pi}{2^m} < 1$ (which is true when $m \ge 2$). Now, set $m=q+3$ to prove the corollary.
\end{proof}


Now, let $p_a$ be the probability of obtaining the basis state $\ket{a}$ on measuring the state $\ket{\psi}$. The amplitude estimation circuit referred to above uses an oracle, denoted $O_a$ to mark the ``good state'' $\ket{a}$, and involves measuring the output of the $AmpEst$ circuit in the standard basis; actually, it suffices to only measure the first register. We can summarise the behaviour of the $AmpEst$ circuit (without the final measurement) in the following lemma.
\begin{lemma}
    Given an oracle $O_x$ that marks $\ket{x}$ in some state $\ket{\psi}$, $AmpEst$ on an input state $\ket{\psi}\ket{0^m}$ generates the following state.
    $$AmpEst \ket{\psi}\ket{0^m} \xrightarrow{} \beta_{x,s}\ket{\psi}\ket{\hat{p_x}} + \beta_{x,\overline{s}}\ket{\psi}\ket{E_x}$$
    where $|\beta_{x,s}|^2$, the probability of obtaining the good estimate, is at least $\frac{8}{\pi^2}$, and $\ket{\hat{p}_x}$ is an $m$-qubit normalized state of the form $\ket{\hat{p}_x} = \gamma_{+}\ket{\hat{p}_{x,+}} + \gamma_{-}\ket{\hat{p}_{x,-}}$ such that for $p\in \{\hat{p}_{x,+}, \hat{p}_{x,-}\} = \pmset_{p_x}~(say)$, $\sin^2(\pi\frac{p}{2^m})$ approximates $p_x$ up to $m-3$ bits of accuracy. Further, $\ket{E_x}$ is an $m$-qubit error state (normalized) such that any basis state in $\ket{E_x}$ corresponds to a bad estimate, i.e., we can express it as $\displaystyle\ket{E_x} = \sum_{t \in \{0,1\}^m}^{t\notin \pmset_{p_x}}\gamma_{t,x}\ket{t}$ in which $|\sin^2\left(\pi\tfrac{t}{2^m}\right) - p_x| > \tfrac{1}{2^{m-3}}$ for any $t \not\in \pmset_{p_x}$.
\end{lemma}

In an alternate setting where the oracle $O_x$ is not provided, $AmpEst$ can still be performed if the basis state $\ket{a}$ is provided --- one can construct a quantum circuit, say $EQ$, that takes as input $\ket{\phi}\ket{x}$ and marks the state $\ket{x}$ of the superposition state $\ket{\phi}$ as described in section~\ref{sec:qbound_fmax}.
We name this extended-$AmpEst$ circuit as $\eqae$ which implements the following operation.

$$\eqae\big(\ket{x}\ket{\psi}\ket{0^m}\big) \xrightarrow{} \ket{x}\big(\beta_{x,s}\ket{\psi}\ket{\hat{p}_x} + \beta_{x,\overline{s}}\ket{\psi}\ket{E_x}\big)$$ where the notations are as defined above and the quantum circuit $EQ$ is used wherever the oracle $O_x$ was used in the previous setting.
In such a scenario, since $\eqae$ is a quantum circuit, we could replace the state $\ket{x}$ by a superposition $\sum_x\alpha_x\ket{x}$.
We then obtain the following.
\begin{corollary}
    Given an $EQ$ circuit, the $\eqae$ on an input state $\sum_x\alpha_x\ket{x}\ket{\psi}\ket{0^m}$ outputs a final state of the form
    $$\eqae\Big(\sum_x\alpha_x\ket{x}\ket{\psi}\ket{0^m}\Big) \xrightarrow{} \sum_x\alpha_x\beta_{x,s}\ket{x}\ket{\psi}\ket{\hat{p}_x} + \sum_x\alpha_x\beta_{x,\overline{s}}\ket{x}\ket{\psi}\ket{E_x}.$$
\end{corollary}
Notice that on measuring the first and the third registers of the output, with probability $|\alpha_x\beta_{x,s}|^2 \ge \frac{8}{\pi^2}|\alpha_x|^2$ we would obtain as measurement outcome a pair $\ket{a}\ket{x}$ where $\sin^2(\pi\frac{a}{2^m}) = \tilde{p}$ is within $\pm \frac{1}{2^{m-3}}$ of the probability $p_x$ of observing the basis state $\ket{x}$ when the state $\ket{\psi}$ is measured.
Observe in this setting that the subroutine essentially estimates the amplitude of all the basis states $\ket{x}$. However, with a single measurement we can obtain the information of at most one of the estimates.

\subsection{Amplitude amplification}

The amplitude amplification algorithm (AA) is a generalization of the novel Grover's algorithm.
Given an $n$-qubit algorithm $A$ that outputs the state $\ket{\phi}=\sum_k\alpha_k\ket{k}$ on $\ket{0^n}$ and a set of basis states $G=\{\ket{a}\}$ of interest, the goal of the amplitude amplification algorithm is to amplify the amplitude $\alpha_a$ corresponding to the basis state $\ket{a}$ for all $\ket{a}\in G$ such that the probability that the final measurement output belongs to $G$ is close to 1.
In the most general setting, one is given access to the set $G$ via an oracle $O_G$ that marks all the states $\ket{a}\in G$ in any given state $\ket{\phi}$; i.e., $O_G$ acts as
$$O_G \sum_k\alpha_k\ket{k}\ket{0} \xrightarrow{}\sum_{a\notin G}\alpha_a\ket{a}\ket{0} + \sum_{a\in G}\alpha_a\ket{a}\ket{1}.$$

Now, for any $G$, any state $\ket{\phi} = \sum_k\alpha_k\ket{k}$ can be written as 
$$\ket{\phi} = \sum_k\alpha_k\ket{k} = \sin(\theta)\ket{\nu} + \cos(\theta)\ket{\overline{\nu}}$$ where $\sin(\theta) = \sqrt{\sum_{a\in G}|\alpha_a|^2}$, $\ket{\nu} = \frac{\sum_{a\in G}\alpha_a\ket{a}}{\sqrt{\sum_{a\in G}|\alpha_a|^2}}$ and $\ket{\overline{\nu}} = \frac{\sum_{a\notin G}\alpha_a\ket{a}}{\sqrt{\sum_{a\notin G}|\alpha_a|^2}}$.
Notice that the states $\ket{\nu}$ and $\ket{\overline{\nu}}$ are normalized and are orthogonal to each other.
The action of the amplitude amplification algorithm can then be given as
$$AA\Big(\sum_k\alpha_k\ket{k}\ket{0}\Big) = AA\big(\sin(\theta)\ket{\nu} + \cos(\theta)\ket{\overline{\nu}}\big)\ket{0} \xrightarrow{} \sqrt{(1-\beta)}\ket{\nu}\ket{1} + \sqrt{\beta}\ket{\overline{\nu}}\ket{0}$$ where $\beta$ satisfies $|\beta| < \delta$ and $\delta$ is the desired error probability.
This implies that on measuring the final state of AA, the measurement outcome $\ket{a}$ belongs to $G$ with probability $|1-\beta|$ which is at least $1-\delta$.


	
    

\section{Lower bound for randomised non-linearity testing}
\label{appendix:rlb}

Consider the following theorem from \cite{quantum_random_lower}.

\begin{theorem}[Theorem~2~\cite{quantum_random_lower}]\label{thm:quant_rand_lower}
Let $S$ and $S'$ be two sets and let $f: S\xrightarrow{} S'$ be a function. Consider a weight scheme where
\begin{enumerate}
    \item Every pair $(x,y)\in S\times S$ is assigned a non-negative weight $w(x,y)$ such that $w(x,y)=0$ whenever $f(x)=f(y)$.
    \item Every tuple $(x,y,i)$ is assigned a non-negative weight $w'(x,y,i)$ such that $w'(x,y,i)=0$ whenever $f(x)=f(y)$ or $x_i = y_i$.
\end{enumerate}
Define $wt(x) = \Sigma_{y} w(x,y)$ and $v(x,i) = \Sigma_{y} w'(x,y,i)$ for all $x$ and for all $i$. If $w'(x,y,i)w'(y,x,i) \ge w^2(x,y)$ for all $x,y,i$ such that $x_i\neq y_i$, then 
$$
Q_2(f) = \Omega\Bigg(\min_{\substack{x,y,i \\ w(x,y)\neq 0, x_i\neq y_i}} \sqrt{\frac{wt(x)wt(y)}{v(x,i) v(y,i)}}\Bigg).
$$

Moreover, if  $w'(x,y,i)w'(y,x,i) \ge w(x,y)$ for all $x,y,i$ such that $x_i\neq y_i$, then

$$
R(f) = \Omega\Bigg(\min_{\substack{x,y,i \\ w(x,y)\neq 0, x_i\neq y_i}} \max\Bigg\{\frac{wt(x)}{v(x,i)}, \frac{wt(y)}{v(y,i)}\Bigg\}\Bigg).
$$
where $Q_2(f)$ is the bounded error quantum query complexity and $R(f)$ is the classical randomised query complexity of $f$.
\end{theorem}

Define a weight scheme such that
$$
w(x,y) = \begin{cases} 0, \text{if } f(x)=f(y)\\ 
                       1, \text{else}
         \end{cases}
\text{; }
w'(x,y,i) = \begin{cases} 0, \text{if } f(x)=f(y) \text{ or } x_i = y_i\\ 
                       1, \text{else}
         \end{cases}.
$$
Note that in this weight scheme it is always true that $w'(x,y,i)\cdot w'(y,x,i) \ge w^2(x,y)$ and $w'(x,y,i) \cdot w'(y,x,i) \ge w(x,y)$ for all $x,y,i$ such that $x_i\neq y_i$.
Let $X$ be a non-empty subset of the set of strings $x$ such that $f(x)=0$ and $Y$ be a non-empty subset of the set of strings $y$ such that $f(y)=1$ where $f$ is some decision function which takes inputs of size $n$. Let $[n] = \{1,\cdots,n\}$.
Also let $S = X \cup Y$ and $R = X \times Y$.
Define $m_x$ (resp. $m_y$) for any $x\in X$ (resp. $y\in Y$) as the number of $y\in Y$(resp. $x\in X$) such that $(x,y)\in R$.
Similarly, for any $x\in X$ (resp. $y\in Y$) and $i\in [n]$ define $l_{(x,i)}$ (resp. $l_{(y,i)}$) as the number of $y\in Y$(resp. $x\in X$) such that $(x,y)\in R$ and $x_i\neq y_i$.
From the definition of the weights from Theorem~\ref{thm:quant_rand_lower}, we can see that $wt(x) = m_x$, $wt(y) = m_y$, $v(x,i) = l_{(x.i)}$ and $v(y,i) = l_{(y,i)}$ for any $x\in X$, $y\in Y$ and $i\in [n]$.
Adapting the definition from Theorem~\ref{thm:qa_method}, let $m = \min_x\{m_x\}$, $m' = \min_y\{m_y\}$, $l = \max_{x,i}\{l_{(x,i)}\}$ and $l' = \max_{y,i}\{l_{(y,i)}\}$.
Using these definitions, we can explicitly see the equivalence of the forms of $Q_2(f)$ in Theorem~\ref{thm:qa_method} and Theorem~\ref{thm:quant_rand_lower} as 
$$Q_2(f) = \Omega\Bigg(\min_{\substack{x,y,i \\ w(x,y)\neq 0, x_i\neq y_i}} \sqrt{\frac{wt(x)wt(y)}{v(x,i) v(y,i)}}\Bigg) = \Omega\Big(\sqrt{\frac{m\cdot m'}{l\cdot l'}}\Big).$$

Since $X$ and $Y$ are non-empty and $R = X\times Y$, we know for  any $y$, $m_y\ge 1$.
Next, since for any $x\in X$ and $y\in Y$, we have $f(x)\neq f(y)$, it is obvious that $x\neq y$. So for any $y\in Y$ there is at least one index $i\in [n]$ such that $x_i\neq y_i$ and hence $l_{(y,i)}\ge 1$.
Now, see in the proofs of Theorem~\ref{thm:lb_fhat_decide1} and Theorem~\ref{thm:lb_fhat_decide2} that $m' = l' = 1$.
Hence we have $wt(y) = m_y = 1$ and $v(y,i) = l_{(y,i)} = 1$ for any $y\in Y$ and $i\in [n]$.
From the proofs of Theorem~\ref{thm:lb_fhat_decide1} and Theorem~\ref{thm:lb_fhat_decide2} we also have $\frac{m}{l} = \min_{x,i} \big(\frac{m_x}{l_{(x,i)}}\big) \ge 1$. Hence $\frac{m_x}{l_{(x,i)}} \ge 1$ for any $x\in X$ and $i\in [n]$ and so $\max\Bigg\{\frac{wt(x)}{v(x,i)}, \frac{wt(y)}{v(y,i)}\Bigg\} = \frac{wt(x)}{v(x,i)}$.
This implies that
$$
R(f) = \Omega\Bigg(\min_{\substack{x,y,i \\ w(x,y)\neq 0, x_i\neq y_i}} \frac{wt(x)}{v(x,i)}\Bigg) = \Omega\Big(\frac{m}{l}\Big),
$$
leading us the following theorem.

\begin{theorem}
Any classical randomised algorithm uses $\Omega(\frac{1}{\lambda})$ queries to estimate the non-linearity of a given function with $\lambda$ accuracy.
\end{theorem}

\section{Alternative Quantum Approaches for {\tt BoundFMax}}\label{sec:approach}

Recall that there are three components that contribute to the complexity of {\tt CBoundFMax}. The innermost component is the classical estimation with complexity $\tilde{O}(\frac{1}{\epsilon^2}\log{\frac{1}{\delta}})$. 
This is nested inside the loop over all suffixes of any particular length which contributes $O(\frac{1}{\tau-2\epsilon})$.
This in turn is nested inside an outer loop over all levels of the binary tree which gives a $O(n)$ to the complexity.
Improving the speed of any of the three components will result in an overall speedup.
In the algorithms that follow, we tried to replace the existing classical component with a quantum component which has a speedup over its classical counterpart. 

\subsection{{\tt CBoundFMax-QPWC}}\label{subsec:cboundfmax-qpwc}

The most naive way to introduce ``quantumness'' in {\tt CBoundFMax} is by replacing the classical estimation of PWC in Algorithm~\ref{algo:cbound_fmax} by a quantum estimation of PWC.
The new algorithm which we call {\tt CBoundFmax-QPWC} is given as Algorithm~\ref{algo:cbound_fmax_qpwc}.

    \begin{algorithm}[!h]
	\caption{\label{algo:cbound_fmax_qpwc}Algorithm \texttt{CBoundFMax-QPWC}}
	\begin{algorithmic}
	    \Require threshold $\tau \in (0,1)$, confidence
	    $\epsilon \in (0,\tau)$, error $\delta \in (0,1)$
	    \State Initialise a FIFO list $Q = \{\varepsilon\}$, where $\varepsilon$ denotes the empty string
	    \While{$Q$ is not empty}
		\State remove prefix $\mathfrak{p}$ from $Q$
		\For{suffix $\mathfrak{s}$ from $\{0,1\}$}
		    \State childprefix $\fcp = \mathfrak{p^\frown s}$
		    \State Estimate $e = \sum_{x\text{ with prefix $\fcp$}} \hat{f}(x)^2$ with accuracy $\epsilon$ and error $(\frac{\tau-2\epsilon}{2n})\delta$. \Comment{Lemma~\ref{thm:PWC}} 
		    \If{${e} \ge \tau-\epsilon$}
			\State If $len(\fcp) < n$, add $\mathfrak{cp}$
			to $Q$
			\State Else (i.e., $len(\fcp) = n$), \Return
	    {\tt TRUE}
		    \EndIf
		\EndFor
	    \EndWhile
	    \State \Return {\tt FALSE}
	\end{algorithmic}
    \end{algorithm}

The correctness of the algorithm clearly follows from the correctness of Algorithm~\ref{algo:cbound_fmax}.
Now see that the complexity of the quantum estimation can be obtained as $O\big(\frac{1}{\epsilon}\log(\frac{1}{\delta})\big)$ from Theorem~\ref{thm:PWC}.
The complexity of the classical outer loop is {$O\big(\frac{n}{\tau-2\epsilon}\big)$} as derived in Section~\ref{sec:classical_algo}.
Hence, the query complexity of {\tt CBoundFMax-QPWC} turns out as $O\big(\frac{n}{\epsilon(\tau-2\epsilon)}\log(\frac{1}{\delta})\big)$ queries, a slight improvement over the {\tt CBoundFMax}.

Using {\tt CBoundFMax-QPWC} to solve the {\tt BoundFMax} problem in Algorithm~\ref{algo:intervalsearch}, we obtain a hybrid algorithm of query complexity $O(\frac{n}{\epsilon^2}\log{\frac{1}{\delta}})$ that outputs an interval of length at most $\epsilon$ that contains $\hat{f}_{max}^2$ with probability at least $1 - \delta$, given $\epsilon$ and $\delta$.

\subsection{{\tt QCBoundFMax}}\label{subsec:qcboundfmax}

Now we describe another optimisation that changes one more classical operation to a quantum one. Recall that in the level order traversal of the binary tree, at any particular level, there can be at most $\frac{1}{\tau-2\epsilon}$ many prefixes such that the estimate of the $PWC$ at those points are greater than the threshold.
Since the estimation of $PWC$ at the next level is performed over all points with these prefixes, the number of estimations due to these prefixes is of the order $O(\frac{1}{\tau-2\epsilon})$.
In this version of the algorithm, which we call {\tt QCBoundFMax}, we replace the classical level order traversal with a quantum level order traversal and use superposition and amplitude amplification to our benefit.

Our quantum algorithm {\tt QCBoundFMax} is described in  Algorithm~\ref{algo:qcbound_fmax}. 
We use $Ri$ to represent the $i^{th}$ register used in the quantum circuit corresponding to the algorithm.
In the algorithm, we also use a few smaller circuits as described in Section~\ref{sec:qbound_fmax}.

\begin{algorithm}
    \caption{Algorithm \texttt{QCBoundFMax} \label{algo:qcbound_fmax}}
    \begin{algorithmic}[1]
        \Require Threshold $\tau$, accuracy $\epsilon$ and error $\delta$.
        \State Set $\tau' = \tau - \epsilon$ and $q = \ceil{\log(\frac{1}{\epsilon})}+3$
        \State Set $\tau_1 = \left\lfloor{\frac{2^q}{\pi}\sin^{-1}(\sqrt{\tau'})}\right\rfloor$ if $\sin^{-1}(\sqrt{\tau'}) \le \frac{\pi}{2}$ and $\tau_1 = \ceil{\frac{2^q}{\pi}\sin^{-1}(\sqrt{\tau'})}$ if $\sin^{-1}(\sqrt{\tau'}) > \frac{\pi}{2}$.
        \State Initialise the circuit as $\ket{0^n}\ket{0^n}\ket{0^q}\ket{\tau_1}\ket{0^n}$.
        \State Apply $H^{\otimes n}$ on $R1$.
        \For{$i$ in $\{1,...,n\}$}
        \State Apply quantum amplitude estimation ($AmpEst$) on $DJ$ with $i$ qubits of $R2$ as the input register, $R3$ as the precision register and $R1$ is used to determine the ``good state''. $AmpEst$ is called with error at most $\delta/2n$ and additive accuracy $\frac{1}{2}\cdot \frac{1}{2^q}$.\label{line:amp_est_1}
        \State Use ${\tt HD_q}$ on $R3$ and $R4$ separately.\label{line:half_dist_1}
        \State Use ${\tt CMP}$ on $R3$ and $R4$ as input registers and $R5$ as output register.\label{line:q_compare_1}
        \State Apply Fixed Point Amplitude Amplification (FPAA) $\frac{1}{\sqrt{\tau-2\epsilon}}$ times on the $i^{th}$ qubit of $R5$ with error at most $\delta/2n$ and measure $i^{th}$ qubit of $R5$ as $m_i$.
        \If{$m_i = \ket{0}$}
            \State \Return {\tt FALSE}
        \ElsIf{$m_i = \ket{1}$  and $i = n$}
            \State \Return{\tt TRUE}
        \EndIf
        \State Reset $R3$ to $\ket{0^q}$ and $R4$ to $\ket{\tau_1}$ by applying $HD_q^{\dagger}$ individually on them.
        \EndFor
    \end{algorithmic}
\end{algorithm}

One can see that in {\tt QCBoundFMax} the selection of desired prefixes (i.e., any prefix p such that $PWC(p)\ge \tau$) for the next round is done quite differently from what is used in {\tt CBoundFMax-QPWC}.
While in {\tt CBoundFMax-QPWC} the desired prefixes for the round $i$ are selected by estimating $PWC$ of each of the prefixes selected at round $i-1$ one at a time and then comparing them with the threshold, in {\tt QCBoundFMax} we use a clever combination of amplitude estimation without measurement, amplitude amplification and two constant time subroutines $HD_q$ and $CMP$.
The estimation and comparison using $AmpEst$, $HD_q$ and $CMP$ is exactly as explained in Section~\ref{sec:qbound_fmax} except that here the estimate is of the amplitude of the prefixes $p$ that were selected in round $i$.
Also of difference is that the {\tt EQ} circuit in the amplitude estimation module here works such that in the $i^{th}$ round, {\tt EQ} maps the $2n$-qubit basis states $\ket{x,y}$ to $(-1)\ket{x,y}$ if $x_j = y_j$ for all $j\in\{0,1,\cdots, i-1\}$ where $x_j$ denotes the $j^{th}$ bit of the string $x$ and similarly for $y_j$.
Post estimation and comparison, the $i^{th}$ output bit of any $i$-bit prefix $p$ contains $\ket{1}$ if $PWC(p)$ is greater than the threshold and $\ket{0}$ otherwise.
We next use amplitude amplification to amplify only those states whose $i^{th}$ output qubit is $\ket{1}$. Since at any level there can be at most $\frac{1}{\tau-2\epsilon}$ many prefixes which are greater than the threshold, we use {\tt FPAA} $O(\frac{1}{\sqrt{\tau-2\epsilon}})$ many times to amplify the desired states.
This process is repeated in loop over all possible lengths of prefixes.
Hence the total query complexity of {\tt QCBoundFMax} is $O(n)\times O(\frac{1}{\sqrt{\tau-2\epsilon}})\times O(1) \times O(\frac{1}{\epsilon}) = O(\frac{n}{\epsilon\sqrt{(\tau-2\epsilon)}})$ queries.
Notice that this is an $O(\frac{1}{\sqrt{\tau-2\epsilon}})$ speedup over {\tt CBoundFMax-QPWC}. 

It easily follows that the use of {\tt QCBoundFMax} { to solve the {\tt BoundFMax} problem} in Algorithm~\ref{algo:intervalsearch} gives us a quantum algorithm that given additive accuracy $\epsilon$ and error probability $\delta$ outputs an interval of length at most $\epsilon$ that contains $\hat{f}_{max}^2$ with probability $1-\delta$ using $O(\frac{n}{\epsilon^{3/2}} \log{\frac{1}{\delta}})$ queries to the oracle of the function.

Finally, observe that by cleverly setting the initial state and ignoring the first $n-1$ rounds in $\tt QCBoundFMax$ algorithm we obtain Algorithm~\ref{algo:qbound_fmax} with a better query complexity.

\section{Subroutines used in {\tt QB\lowercase{ound}FM\lowercase{ax}} Algorithm}\label{appendix:subroutine}
In this section we present the algorithms of the $HD_q$, $CMP$ and ${\tt Cond\text{-}MAJ}_y$ subroutines used in Algorithm~\ref{algo:qbound_fmax}. The string representations of the integers are indexed in the reverse order, i.e., the bit-representation of an integer $z$ is represented as $z=z_0 z_1 z_2 \ldots$ where $z_0$ denotes the most significant bit of $z$. We have used the following quantum operations in those algorithms.

\begin{description}
    \item[{\tt CX(a,b)} :] Controlled-NOT gate where {\tt X} gate is applied on qubit {\tt b} if the control qubit {\tt a} is in the state $\ket{1}$.
    \item[${\tt \overline{C}X(a,b)}$ :] Controlled-NOT gate where {\tt X} gate is applied on qubit {\tt b} if the control qubit {\tt a} is in the state $\ket{0}$.
    \item[${\tt C\overline{C}^{i}X(a,b_1,b_2, \cdots, b_i, c)}$ :] Multi-controlled NOT gate where {\tt X} gate is applied on qubit {\tt c} if the first control qubit {\tt a} is in the state $\ket{1}$ and the next $i$ control qubits ${\tt b_1,b_2, \cdots, b_i}$ are in the state $\ket{0}$.
    \item[{\tt SWAP(A,B)} :] Swap gate that swaps register $A$ with register $B$ given that the registers $A$ and $B$ are of the same size.
\end{description}

\subsection{Quantum circuit for $HD_q$ operation}
\begin{figure}
    \centering
    \includegraphics[width=\linewidth]{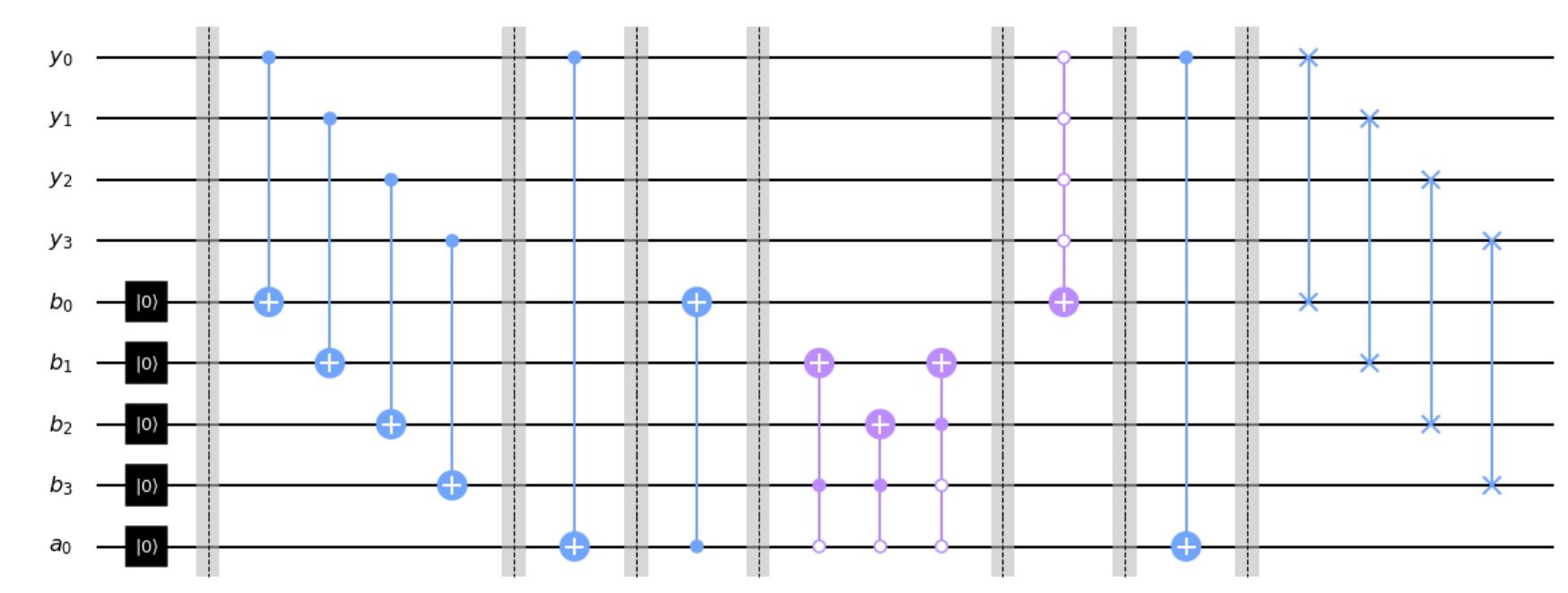}
    \caption{The ${\tt HD_q}$ circuit for a 4 qubit input $y$ (Created using Qiskit).}
    \label{fig:hd}
\end{figure}

The ${\tt HD_q}$ operation was defined as $\ket{y}\ket{b} \mapsto \ket{b\oplus\tilde{y}}\ket{y}$ where $y,b\in \{0,1\}^q$ and $\tilde{y}$ is the bit string corresponding to the integer $\abs{2^{q-1} - y_{int}}$.
An algorithm to implement this operator is described in Algorithm~\ref{algo:HD}. We also show the quantum circuit corresponding to ${\tt HD_4}$ in Figure~\ref{fig:hd}. Note that the register $a$ is an ancilla register.

\begin{algorithm}
    \caption{Subroutine \texttt{$HD_q$} \label{algo:HD}}
    \begin{algorithmic}[1]
        \State Initialise ${\tt R_1 R_2 R_3} = \ket{y}\ket{0^n}\ket{0}$.
        \For{i in $\{0,1,2,\cdots, n-1\}$}
            \State Apply ${\tt CX(R_1[i],R_2[i])}$.
        \EndFor
        \State Apply ${\tt CX(R_1[0],R_3)}$.
        \State Apply ${\tt CX(R_3, R_2[0])}$.
        \State Initialise an empty array $S$.
        \For{i in $\{0,1,2,\cdots, n-1\}$}
            \State Append $R_2[n-i-1]$ to $S$.
            \For{k in $\{1,n-i-2\}$}
                \State Apply ${\tt \overline{C}^{i+1}CX(R_3, S[0],S[1],\cdots, S[i], R_2[k])}$.
            \EndFor
        \EndFor
        \State Apply ${\tt \overline{C}^{n}X(R_1[0],R_1[1], \cdots, R_1[n-1],R_2[0]) }$.
        \State Apply ${\tt CX(R_1[0], R_3)}$.
        \State Apply ${\tt SWAP(R_1, R_2)}$.
    \end{algorithmic}
\end{algorithm}

The algorithm behind ${\tt HD_q}$ is based on bitwise manipulations. We use $y$ to denote the input integer ($0 \le y < 2^q$), available in the first register. We will focus on obtaining $\ket{|y - 2^{q-1}|}$ in a register. The simple case is $y \ge 2^{q-1}$ when we need to simply set the most significant bit of $y$ to 0~\footnote{For instance if $y = 111$, then $\tilde{y}=100 - 111 = 011$.}. For the other case, i.e., $y < 2^{q-1}$, we have to subtract $y$ (available in binary) from $2^{q-1}$ (1 followed by $q-1$ zeroes). The algorithm implements this by first finding out the right{most} bit of $y$ that is set to $1$, denoted $t$; all bits to the right of $t$ are 0 and do not contribute to the difference. The $t$-th bit of the difference is set to 1. And all the bits to the left of $t$ are flipped (they are subtracted from 1 created by the cascading borrowing effect when the $t$-bit of the difference was calculated)~\footnote{For instance, if we have $y = 0110$, then we need $10000-00110$. The first bit from the right that is set to one is the second least significant bit. Then all bits between the second least significant bit and the most significant bit is flipped. So we have $10000-00110 = 01010$.}. And of course, if all the bits of $y$ are 0, then we just flip the first bit of the answer.



\subsection{Quantum circuit for $CMP$ operation}
\begin{figure}
    \centering
    \includegraphics[width=\linewidth]{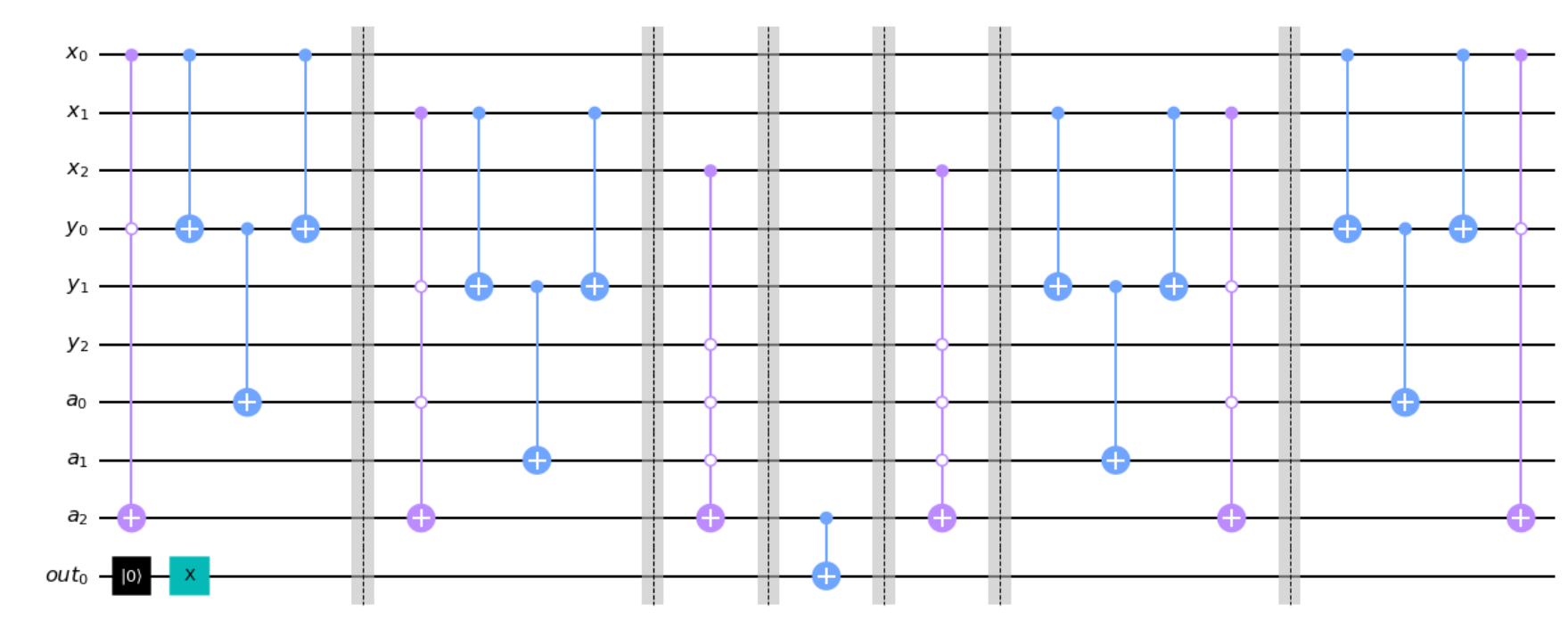}
    \caption{The {\tt CMP} circuit of the inputs $y$ and $t$ of size $3$ (Created using Qiskit).}
    \label{fig:cmp}
\end{figure}

Next, we present the algorithm for the subroutine ${\tt CMP}$ in Algorithm~\ref{algo:CMP}.
Figure~\ref{fig:cmp} contains the circuit for the {\tt CMP} subroutine with $x$ as the first input and $y$ as the second input. 
It is to be noted that the register $a$ in the circuit is an ancilla register. The state of the register $a$ before and after the subroutine is applied is $\ket{0^{n}}$.
The $out$ register contains the result of the comparison in the form $\ket{x\le y}$.

\begin{algorithm}
    \caption{Subroutine \texttt{CMP} \label{algo:CMP}}
    \begin{algorithmic}[1]
        \State Initialise ${\tt R_1 R_2 R_3 R_4} = \ket{x}\ket{y}\ket{0^{n}}\ket{0}$.
        \State Apply ${\tt X(R_4)}$.
        \State Initialise an empty array $A$.
        \For{$i$ in $\{0, 1, \cdots, n-1\}$}\label{line:4}
            \State Apply ${\tt C\overline{C}^{i+1}X(R_1[i], R_2[i], A[0],A[1],\cdots,A[i-1], R_3[n-1])}$.
            \If{$i \neq n-1$}
                \State Apply ${\tt CX(R_1[i],R_2[i])}$.
                \State Apply ${\tt CX(R_2[i],R_3[i])}$.
                \State Apply ${\tt CX(R_1[i],R_2[i])}$.
            \EndIf
            \State Append ${\tt R_3[i]}$ to array $A$.
        \EndFor\label{line:12}
        \State Apply ${\tt CX(R_3[n-1],R_4)}$.
        \State Apply the conjugate transpose of all the operations from line~\ref{line:4} to line~\ref{line:12}.
    \end{algorithmic}
\end{algorithm}

The algorithm applies the standard method of comparing two bit-strings. It compares the $k$-th bit of $x$ and $y$, starting from the most significant bit and moving right, until it finds some $k$ such that $x_k \not= y_k$. For this position, the answer bit is flipped (i.e., we claim that $x > y$) if $x_k=1, y_k=0$; otherwise we claim that $x \le y$. We also start the answer qubit in $\ket{1}$, so if $x \le y$, then the answer qubit remains in the state $\ket{1}$.

\newpage
\subsection{Quantum circuit for ${Cond\text{-}MAJ}_y$ operation}
Let $X_1 \ldots X_k$ be Bernoulli random variables with success probability $p > 1/2$. Let $Maj$ denote their majority value (that appears more than $k/2$ times). A common result that is often used, and can be proved easily using Chernoff's bound \footnote{$\Pr[\sum X_i > \frac{k}{2}] \ge 1 - \exp(-\frac{1}{2p}n(p-\frac{1}{2})^2)$}, is that $Maj$ has a success probability at least $1-\delta$, for any given $\delta$, if we choose $k \ge \tfrac{2p}{(p-1/2)^2} \ln \tfrac{1}{\delta}$. We require a quantum formulation of the same.

Suppose we have k copies of the quantum state $\ket{\phi} = \sum_{x}\alpha_x\ket{x}\Big(\ket{\psi_{0,x}} \ket{0} + \ket{\psi_{1,x}} \ket{1}\Big)$ in which we define ``success'' as observing $\ket{1}$ in the last register (without loss of generality) of $\ket{\phi}$ given that the first register is $\ket{y}$.
Let $p = \big\| \ket{\psi_{1,y}} \big\|^2$ denote the probability of success and let $p > 1/2$. 
Now, define the conditional majority ${\tt Cond\text{-}MAJ}_y$ as the operator $ \mathbb{U}_y$ that performs the $MAJ$ operator on the third registers of each copy of $\ket{\phi}$ if the corresponding first register in that copy is $\ket{y}$ and stores the output in the second register of an answer state if the first register of the answer state is $\ket{y}$.
Suppose we apply $\mathbb{U}_y$ on $\ket{\phi}^{\otimes}\sum_x \hat{f}(x)\ket{x}\ket{0}$ and we measure the answer register after applying $\mathbb{U}_y$.
Then it is easy to show, essentially using the same analysis as above, that 
$$\mathbb{U}_y\ket{\phi}^{\otimes k}\ket{0} = \alpha_y \ket{y}\Big(\ket{\Gamma_{0,y}} \ket{0} + \ket{\Gamma_{1,y}} \ket{1}\Big) + \Upsilon$$
in which $\big\| \ket{\Gamma_{1,y}} \big\|^2 \ge 1-\delta$ and $\Upsilon = \sum_{x\neq y}\alpha_x\ket{x}\ket{\phi}^{\otimes k}\ket{0}$.

The ${\tt Cond\text{-}MAJ}_y$ operator can be implemented without additional queries and with $poly(k)$ gates and $\log(k)$ qubits.
The algorithm for the implementation is presented in Algorithm~\ref{algo:cond-maj}.

\begin{algorithm}
    \caption{Subroutine $\texttt{Cond-MAJ}_y$ \label{algo:cond-maj}}
    \begin{algorithmic}[1]
        \State Let $t = \lceil \log(k) \rceil$ and $l = \lceil k/2 \rceil$.
        \State Define ${\tt R_i = R_{xi}R_{ci}} = \ket{x_i}\ket{c_i}$ and $n=|y|$.
        \State Initialise ${\tt R_1 \cdots R_k R'_1 \cdots R'_{t}R'' R_a}   = \Big(\bigotimes_{i=1}^{k} \ket{x_i}\ket{c_i}\Big)\ket{0^{\otimes t}}\ket{l-1}\ket{0}$.
        \For{$i$ in $\{1, \cdots, k\}$}\label{line:8-4}
            \For{$j$ in $\{0, \cdots, n-1\}$}
                \State Apply $X(R_{xi}[j])$ if $j^{th}$ bit of $y$ equals $1$.
            \EndFor
            \For{$m \in \{1,\cdots, t\}$}
                \State Apply ${\tt \overline{C}^{n} C^{t+1-m}X(R_{xi},R_{ci},R'_t,R'_{t-1},\cdots, R'_{m})}$.
            \EndFor
        \EndFor \label{line:8-11}
        \State Apply ${\tt CMP(R', R'', R_a)}$ and apply ${\tt X(R_a)}$.
        \State Apply the conjugate transpose of all the operations from line~\ref{line:8-4} to line~\ref{line:8-11}.
    \end{algorithmic}
\end{algorithm}

\section{Proof of Proposition~\ref{prop:tau-rel}}
\label{appendix:prop-proof}

{





\taurelprop*

\begin{proof}


We defined $\tau' = \tau-\epsilon$ and $\tau_1 = \lfloor{\frac{2^q}{\pi}\cdot\sin^{-1}(\sqrt{\tau'})}\rfloor$. Then,
\begin{align*}
    &\tau_1 \le \frac{2^q}{\pi}\cdot\sin^{-1}(\sqrt{\tau'})\\
    &\implies \sin(\frac{\pi}{2^q}\tau_1) \le \sqrt{\tau'}\\
    &\implies \sin^2(\frac{\pi}{2^q}\tau_1) = p_{\tau_1} \le \tau'\\
    &\implies \tau'- p_{\tau_1} \ge 0
\end{align*}

On the other hand, we have,
\begin{align*}
    \tau_1 &> \lfloor{(2^q/\pi)\cdot\sin^{-1}(\sqrt{\tau'})}\rfloor - 1\\
    \implies \sin(\frac{\pi}{2^q}\tau_1) &> \sin(\sin^{-1}(\sqrt{\tau'})-\frac{\pi}{2^q})\\
    &= \sqrt{\tau'}\cos(\frac{\pi}{2^q}) - \cos(sin^{-1}(\sqrt{\tau'}))\sin(\frac{\pi}{2^q})\\
    &=\sqrt{\tau'}\cos(\frac{\pi}{2^q}) - (\sqrt{1-\tau'})\sin(\frac{\pi}{2^q})~~(\text{Uses }\cos(\sin^{-1}(x)) = 1-x^2)\\
    \implies p_{\tau_1} &= \sin^2(\frac{\pi \tau_1}{2^q}) \\
    & \ge \tau'\cos^2(\frac{\pi}{2^q}) + (1-\tau')\sin^2(\frac{\pi}{2^q}) - 2 \sqrt{\tau'}\sqrt{1-\tau'}\cos(\frac{\pi}{2^q})\sin(\frac{\pi}{2^q})\\
    &\ge \tau'\cos^2(\frac{\pi}{2^q}) - (1-\tau')\sin^2(\frac{\pi}{2^q}) - 2 \sqrt{\tau'}\sqrt{1-\tau'}\cos(\frac{\pi}{2^q})\sin(\frac{\pi}{2^q})\\
    &= \tau' - \sin^2(\frac{\pi}{2^q}) - \sqrt{\tau'}\sqrt{1-\tau'}\sin(\frac{2\pi}{2^q})
\end{align*}

Now, notice that,
\begin{align*}
    &\sin^2(\frac{\pi}{2^q}) + \sqrt{\tau'}\sqrt{1-\tau'}\sin(\frac{2\pi}{2^q})\\
    &\le \sin^2(\frac{\pi}{2^q}) + \frac{1}{2}\sin(\frac{2\pi}{2^q})~~(\text{Uses }x(1-x)\le 1/4)\\
    &= \sin^2(\frac{\pi}{2^q}) + \sin(\frac{\pi}{2^q})\cos(\frac{\pi}{2^q})\\
    &= \sin(\frac{\pi}{2^q})(\sin(\frac{\pi}{2^q})+\cos(\frac{\pi}{2^q}))\\
    &\le \frac{2\pi}{2^q}
\end{align*}
The last inequality is due to the fact that $\sin\theta\le \theta$, $\sin(\theta)\le 1$ and $\cos(\theta)\le 1$.
This gives us the following:
\begin{align*}
    p_{\tau_1} &\ge \tau' - \sin^2(\frac{\pi}{2^q}) - \sqrt{\tau'}\sqrt{1-\tau'}\sin(\frac{2\pi}{2^q})\\
    & > \tau' - \frac{2\pi}{2^q}
\end{align*}

So we have $$0 \le \tau' - p_{\tau_1} \le \frac{2\pi}{2^q}.$$

\end{proof}
}

\end{document}